\documentclass[10pt,aps,prb,twocolumn,floatfix,floats,showpacs,superscriptaddress,raggedbottom]{revtex4-1}
\pdfoutput=1
\usepackage{graphicx,latexsym}
\usepackage{dcolumn}
\usepackage{amssymb,amsmath,bm}
\usepackage{subfigure}

\newcommand{\brames}[1]{( #1|}
\newcommand{\ketmes}[1]{|#1)}

\usepackage{hyperref}
\hypersetup{
    pdfnewwindow=true,       
    colorlinks=true,         
    linkcolor=blue,          
    citecolor=blue,          
    filecolor=magenta,       
    urlcolor=black           
}

\def\eq#1{Eq.\ (\ref{#1})}
\def\mb#1{\mbox{\boldmath$#1$}}
\def\fig#1{Fig.\ \ref{#1}}

\begin{document}

\title{Magnetic-field influenced non-equilibrium transport through a quantum
ring \\
with correlated electrons in a photon cavity}

\author{Thorsten Arnold}
\email{tla1@hi.is}
\affiliation{Science Institute, University of Iceland, Dunhaga 3,
        IS-107 Reykjavik, Iceland}
\author{Chi-Shung Tang}
\email{cstang@nuu.edu.tw}
\affiliation{Department of Mechanical Engineering,
        National United University,
        1, Lienda, Miaoli 36003, Taiwan}
\author{Andrei Manolescu}
\affiliation{Reykjavik University, School of Science and Engineering,
        Menntavegur 1, IS-101 Reykjavik, Iceland}
\author{Vidar Gudmundsson}
\email{vidar@hi.is}
\affiliation{Science Institute, University of Iceland, Dunhaga 3,
        IS-107 Reykjavik, Iceland}
%

\begin{abstract}
\ We investigate magnetic-field influenced
time-dependent transport of Coulomb interacting electrons through a
two-dimensional quantum ring in an electromagnetic cavity under
non-equilibrium conditions described by a time-convolutionless
non-Markovian master equation formalism. We take into account the
full electromagnetic interaction of electrons and cavity photons
without resorting to the rotating wave approximation or reduction to
two levels. A bias voltage is applied to semi-infinite leads
along the x-axis, which are connected to the quantum ring.
The magnetic field is tunable to manipulate the
time-dependent electron transport coupled to a photon field with
either x- or y-polarization.
We find that the lead-system-lead current is strongly
suppressed by the y-polarized photon field at magnetic field with
two flux quanta due to a degeneracy of the many-body energy spectrum
of the mostly occupied states. Furthermore, the
current can be significantly enhanced by the y-polarized field at
magnetic field with half integer flux quanta.
\end{abstract}



\maketitle
\section{introduction}
Quantum interference phenomena are essential when developing quantum
devices. Quantum confined geometries conceived for such studies may
consist of which-path
interferometers,~\cite{Buks1998,Sprinzak84:5820} coupled quantum
wires,~\cite{Bertoni2000,Gudmundsson2006} side-coupled quantum
dots,~\cite{Kobayashi04:035319,Valsson2008} or quantum
rings.~\cite{Szafran05:165301,Buchholz2010} These coupled quantum
systems have captured interest due to their potential
applications in electronic spectroscopy tools~\cite{Eugster1991} and
quantum information processing.~\cite{Schroer2006}  Furthermore, the
magnetic flux through the ring system can drive persistent
currents~\cite{PhysRevB.37.6050} and lead to the topological quantum
interference phenomenon known as the Aharonov-Bohm (AB)
effect.~\cite{PhysRev.115.485,Buttiker30.1982,Webb.54.2696,Tonomura1986,Pichugin56:9662}
Both, the persistent current and ring conductance show characteristic oscillations
with period of one flux quantum, $\Phi_0=hc/e$.
Varying either the magnetic field 
or the electrostatic confining potentials 
allows the quantum interference
to be tuned.~\cite{Fuhrer73:205326}

There has been considerable interest in the study of electronic
transport through a quantum system in a strong system-lead coupling
regime driven by periodic time-dependent
potentials,\cite{Tang96:4838,Zeb08:165420,Kienle09:026601,Tang182:65}
longitudinally polarized
fields,\cite{Tang99:1830,Zhou17:6663,Jung85:023420} or transversely
polarized fields.\cite{Tang00:127,Zhou68:155309} On the other hand,
quantum transport driven by a transient time-dependent potential
enables development of switchable quantum devices, in which the
interplay of the electronic system with external perturbation plays
an important
role.\cite{Myohanen09:115107,Stefanucci10:115446,Tahir10:195444,Chen11:115439}
These systems are usually operated in the weak system-lead coupling
regime and described within the wide-band or the Markovian
approximation.\cite{Gurvitz96:15932,Kampen01:00,Harbola06:235309}
Within this approximation, the energy dependence of the electron
tunneling rate or the memory effect in the system are neglected by
assuming that the correlation time of the electrons in the leads is
much shorter than the typical response time of the central system.
However, the transient transport is intrinsically linked to the
coherence and relaxation dynamics and cannot generally be described
in the Markovian approximation.  The energy-dependent spectral
density in the leads has to be included for accurate numerical
calculation.

In order to explicitly explore the transport dynamics with transient
system-lead coupling and electron-photon coupling, a non-Markovian
density-matrix formalism involving the energy-dependent coupled
elements should be considered based on the generalized master
equation
(GME).\cite{Braggio06:026805,Emary07:161404,Bednorz08:206803,Moldoveanu09:073019}
How to appropriately describe the carrier dynamics under
non-equilibrium conditions with realistic device geometries is a
challenging problem.\cite{Gudmundsson09:113007,Gudmundsson10:205319}
More recently, manipulation of electron-photon coupled quantum
systems embedded in an electromagnetic cavity has become one of the most promising
applications in quantum information processing
devices.  Utilizing the giant dipole moments of inter-subband
transitions in quantum wells\cite{Helm00:01,Gabbay11:203103} enables
researchers to reach the ultrastrong electron-photon coupling
regime.\cite{Ciuti05:115303,Devoret07:767,Abdumalikov08:180502}  In
this regime, the dynamical electron-photon coupling mechanism has to
be explored beyond the wide-band and rotating-wave
approximations.\cite{Zela97:106,Sornborger04:052315,Irish07:173601}
Nevertheless, time-dependent transport of Coulomb interacting
electrons through a topologically nontrivial broad ring geometry in
an electromagnetic cavity with quantized photon modes remains
unexplored beyond the Markovian approximation.

In the present work, we explore the transient effects of electronic
transport through a broad quantum ring in a linearly polarized
electromagnetic cavity coupled to electrically biased
leads. This electron-photon coupled system under investigation can
be manipulated by tuning the applied magnetic field and the
polarization of the photon field.  A time-convolutionless (TCL)
version of the GME is utilized to project the time evolution onto
the central system by taking trace with respect to the operators in
the leads.\cite{Davies:BK1976,Spohn53:569,PhysRevA.59.1633}  We
demonstrate the transient transport properties by showing the
many-body (MB) energy spectra, the time-dependence of the electric charge and
current, the magnetic-field dependence of the total charge current with (w) or
without (w/o) photon cavity, the charge density distribution,
the normalized current density distribution and the local current 
coming from an occupation redistribution of the MB states 
in the central quantum ring system.

The paper is organized as follows. In Sec.\ II, the theoretical
model is described. The electron system is embedded in an
electromagnetic cavity by coupling a many-level electron system with
photons using the full photon energy spectrum of a single cavity
mode. In Sec.\ III, we show the numerical results for the dynamical
transient transport properties for different magnetic field and
photon field polarization. Concluding remarks will be presented in
Sec.\ IV.

\section{Model and Theory}

In this section, we describe the central system potential $V_S$
for the broad quantum ring and its connection to the leads.
The electronic ring system is embedded in an electromagnetic cavity by
coupling a many-level electron system with photons using the full
photon energy spectrum of a single cavity mode. The central ring
system is described by a MB system Hamiltonian
$\hat{H}_{S}$ with a uniform perpendicular magnetic
field, in which the electron-electron interaction and the
electron-photon coupling to the x- or y-polarized photon field is
explicitly taken into account. We employ the TCL-GME approach to
explore the non-equilibrium electronic transport when the system is
coupled to leads by a transient switching potential.

\subsection{Quantum ring connected to leads}

The system under investigation is a broad quantum ring connected to
left and right leads $l\in\{L,R\}$ with identical parabolic confining
potentials
\begin{equation}
 V_{l}(\mathbf{r})=\frac{1}{2} m^{*} \Omega_{0}^2 y^2,
\end{equation}
in which the characteristic energy of the confinement is
$\hbar\Omega_0 = 1.0$~meV and $m^* = 0.067 m_e$ is the effective
mass of an electron in GaAs-based material.

\begin{figure}[htbq] 
       \includegraphics[width=0.45\textwidth,angle=0, viewport=34 40 396 260, clip]{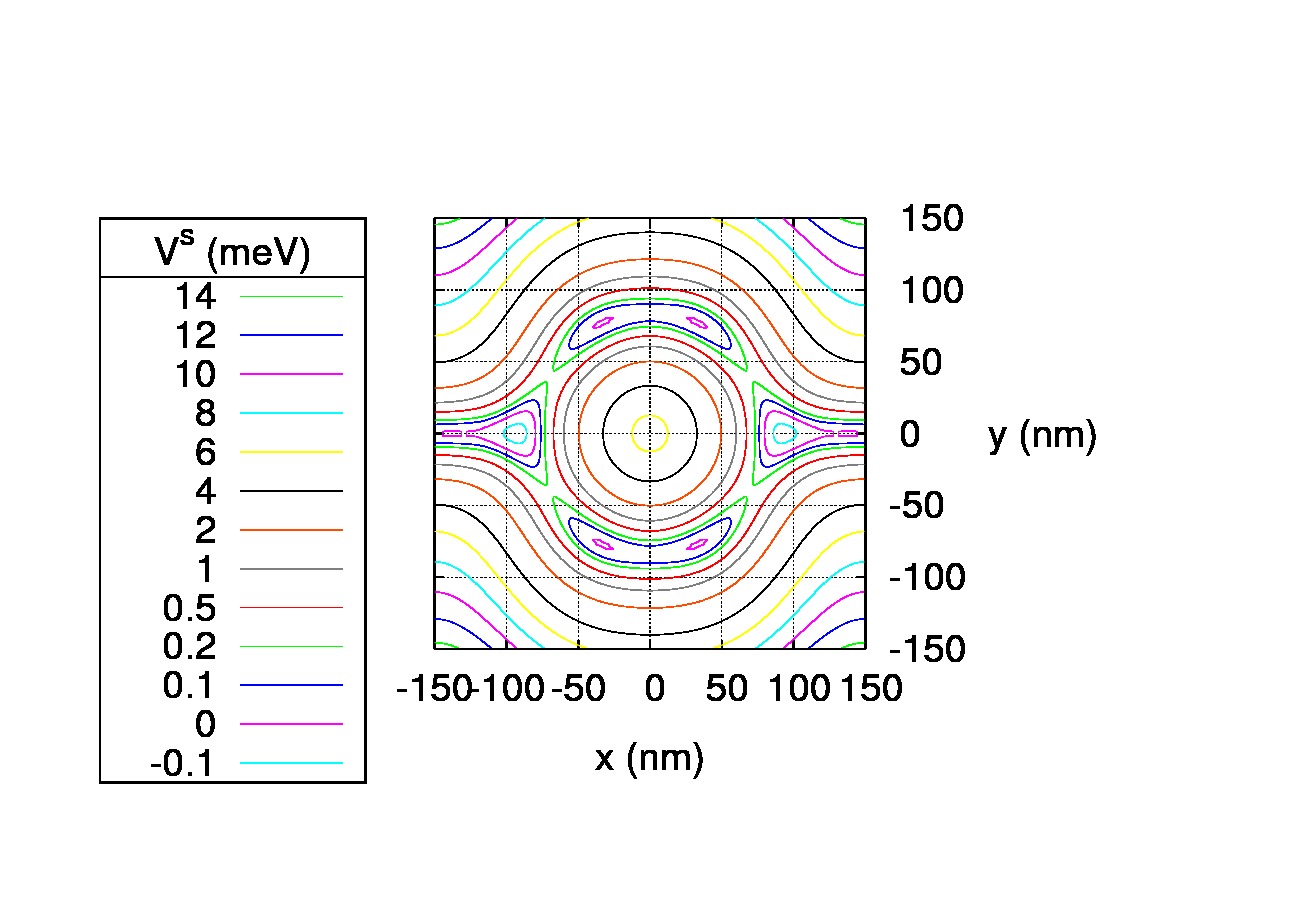}
       \caption{(Color online) Equipotential lines in the central ring  system connected
       to the left and right leads.
       Note that the isolines are refined close to the bottom of the ring structure.}
       \label{ext_pot}
\end{figure}

The quantum ring is embedded in the central system of length $L_x = 300$~\textrm{nm}
situated between two contact areas that will be coupled to the
external leads, as is depicted in \fig{ext_pot}. The system
potential is described by
\begin{eqnarray}
 V_S(\mathbf{r})&=& \sum_{i=1}^{6}V_{i}\exp
 \left[
 -\left(\beta_{xi}(x-x_0)\right)^2 - \left(\beta_{yi}y\right)^2
 \right] \nonumber \\
 &&+\frac{1}{2}m^* \Omega_{0}^2y^2,
\end{eqnarray}
with parameters from table \ref{table:ringpot}.

\begin{table}
 \caption{Parameters of the central region ring potential.}
 \centering
 \begin{tabular}{c    |    c    |    c    |    c    |    c}
\hline\hline
\\ [-1.4ex]
$i$ &  $V_{i}$ in meV &  $\beta_{xi}$ in $\frac{1}{\mathrm{nm}}$
    &  $x_0$ in nm &  $\beta_{yi}$ in $\frac{1}{\mathrm{nm}}$ \\ [0.5ex]
\hline
1 & 9.6 & 0.014 & 150 & 0 \\
2 & 9.6 & 0.014 & -150 & 0 \\
3 & 11.1 & 0.0165 & 0 & 0.0165 \\
4 & -4.7 & 0.02 & 149 & 0.02 \\
5 & -4.7 & 0.02 & -149 & 0.02 \\
6 & -4.924 & 0 & 0 & 0 \\[0.5ex]
\hline\hline
\end{tabular}
\label{table:ringpot}
\end{table}

\subsection{Central system Hamiltonian}

The time-evolution of the closed system with respect to $t=0$
\begin{equation}
 \hat{U}_{S}(t)=\exp\left(-\frac{i}{\hbar}\hat{H}_{S}t\right)
\end{equation}
is governed by the MB system Hamiltonian~\cite{Gudmundsson85:075306}
\begin{eqnarray}
 \hat{H}_{S}&=&\int d^2r\;\hat{\psi}^{\dagger}(\mathbf{r})\left[\frac{1}{2m^{*}}\left(\frac{\hbar}{i}\nabla+
\frac{e}{c}\left[\mathbf{A}(\mathbf{r}) +\hat{\mathbf{A}}^{\mathrm{ph}}(\mathbf{r})\right] \right)^2 \right. \nonumber \\
&& \left. +V_{S}(\mathbf{r}) \right]\hat{\psi}(\mathbf{r})+
\hat{H}_{ee} +\hbar \omega \hat{a}^{\dagger}\hat{a} \, .\label{H_S}
\end{eqnarray}
The first term includes a constant magnetic field $\mathbf{B} = B
\hat{\mb{z}}$, in Landau gauge being represented by $
\mathbf{A}(\mathbf{r})=
 -By \hat{\mb{x}}$. The second term is the exactly treated electron-electron interaction
\begin{equation}
 \hat{H}_{ee}=\int d^2r \int d^2 r' \hat{\psi}^{\dagger}(\mathbf{r})\hat{\psi}^{\dagger}
 (\mathbf{r'}) V_{ee}(\mathbf{r},\mathbf{r'})
\hat{\psi}(\mathbf{r'})\hat{\psi}(\mathbf{r})\, ,
\end{equation}
where
\begin{equation}
 V_{ee}(\mathbf{r},\mathbf{r'})=\frac{e^2}{2\kappa \sqrt{|\mathbf{r}-\mathbf{r'}|^2+\eta^2}}
\end{equation}
with $e>0$ being the magnitude of the electron charge and
$\eta=1.0\times 10^{-15}$~nm being a numerical regularization parameter.  In
addition, the last term in \eq{H_S} indicates the quantized photon field,
where $\hat{a}$ and $\hat{a}^{\dagger}$ are the
photon annihilation and creation operators, respectively,
and $\hbar\omega$ is the photon excitation energy. The photon field
interacts with the electron system via the vector potential
\begin{equation}
 \hat{\mathbf{A}}^{\mathrm{ph}}(\mathbf{r})=A(\hat{a}+\hat{a}^{\dagger})
 \begin{cases} \mathbf{e}_x, & \mathrm{TE}_{011} \\ \mathbf{e}_y, & \mathrm{TE}_{101} \end{cases}
\end{equation}
for longitudinally-polarized (x-polarized) photon
field ($\mathrm{TE}_{011}$) or transversely-polarized (y-polarized)
photon field ($\mathrm{TE}_{101}$). The electron-photon coupling
constant $g^{EM}=eA a_w \Omega_w/c$ scales with the amplitude $A$ of
the electromagnetic field. For reasons of comparison, we also
consider results without photons in the system. In this case,
$\hat{\mathbf{A}}^{\mathrm{ph}}(\mathbf{r})$ and $\hbar \omega
\hat{a}^{\dagger}\hat{a}$ drop out from the MB system Hamiltonian in
\eq{H_S}.

\subsection{Time-convolutionless generalized master equation approach}

The TCL-GME~\cite{PhysRevA.59.1633} is an alternative non-Markovian
master equation to the Nakajima-Zwanzig (NZ)
equation,\cite{Nakajima, Zwanzig, 1367-2630-11-11-113007,
PhysRevB.81.155442} which is {\em local} in time. We assume, the
initial total statistical density matrix can be written as 
a product of the system and leads density matrices, before
switching on the coupling to the leads,
\begin{equation}
 \hat{W}(0)=\hat{\rho}_L \otimes \hat{\rho}_R \otimes \hat{\rho}_S(0),
\end{equation}
with $\rho_l$, $l\in \{L,R\}$, being the normalized density matrices
of the leads. The coupling Hamiltonian between the central system
and the leads reads
\begin{equation}
 \hat{H}_{T}(t)= \sum_{l=L,R}\int dq\;\chi^{l}(t)\left[\hat{\mathfrak{T}}^{l}(q)\hat{C}_{ql}
 +\hat{C}_{ql}^{\dagger} \hat{\mathfrak{T}}^{l\dagger}(q) \right]\,
 .
\end{equation}
Here, $\hat{C}_{ql}^{\dagger}$ is the electron creation operator for state $q$ and lead
$l$ and
\begin{equation}
 \hat{\mathfrak{T}}^{l}(q)
 = \sum_{\alpha \beta}\ketmes{\alpha}\brames{\beta}\sum_{a}T_{qa}^{l}
 \brames{\alpha} \hat{C}_{a}^{\dagger}\ketmes{\beta}
 \label{mathfrakT}
\end{equation}
with the creation operator, $\hat{C}_{a}^{\dagger}$,
for the single-electron state (SES) $a$ in the central system, i.e. 
the eigenstate $a$ of the first term of \eq{H_S} with $\hat{\mathbf{A}}^{\mathrm{ph}}(\mathbf{r})=0$.
The coupling is switched on at $t=0$ via the switching function
\begin{equation}
 \chi^{l}(t)=1-\frac{2}{e^{\alpha^{l}t}+1}
\end{equation}
with switching parameter $\alpha^l$. \eq{mathfrakT} is written in
the system Hamiltonian MB eigenbasis $\{\ketmes{\alpha}\}$.  The
coupling tensor \cite{Gudmundsson85:075306}
\begin{equation}
 T_{qa}^{l}=\int_{\Omega_S^l} d^2r\; \int_{\Omega_l} d^2r'\;
 \psi_{ql}^{*}(\mathbf{r'}) g_{aq}^{l}(\mathbf{r},\mathbf{r'})\psi_a^S(\mathbf{r})
\end{equation}
couples the extended lead SES
$\{\psi_{ql}(\mathbf{r})\}$ with energy spectrum $\{\epsilon^l(q)\}$
to the system SES $\{\psi_a^S(\mathbf{r})\}$ with energy spectrum
$\{E_a\}$ that reach into the contact regions,
\cite{1367-2630-11-11-113007} $\Omega_S^l$ and $\Omega_l$, of system
and lead $l$, respectively, and
\begin{eqnarray}
 g_{aq}^l(\mathbf{r},\mathbf{r'})&=&g_0^l\exp\left[-\delta_x^l(x-x')^2-\delta_y^l(y-y')^2 \right]\nonumber \\
&&\times \exp{
\left(
 -\frac{|E_a-\epsilon^l(q)|}{\Delta^l_E}
\right)}\, .
\end{eqnarray}
Here, $g_0^l$ is the lead coupling strength. In addition, $\delta^l_x$
and $\delta^l_y$ are the contact region parameters for lead $l$ in x-
and y-direction, respectively. Moreover, $\Delta^l_E$ denotes the
affinity constant between the central system SES energy levels $\{E_a\}$
and the lead energy levels $\{\epsilon^l(q)\}$.

In this work, we derive the TCL-GME~\cite{PhysRevA.59.1633} in the
Schr\"{o}dinger picture. In this picture, the reduced density
operator (RDO) of the system,
\begin{equation}
 \hat{\rho}_S(t)= \mathrm{Tr}_{L}\mathrm{Tr}_{R}[\hat{W}(t)],
\end{equation}
evolves to second order in the lead coupling strength in time via
\begin{eqnarray}
\dot{\hat{\rho}}_{S}(t)&=&-\frac{i}{\hbar}[\hat{H}_{S},\hat{\rho}_{S}(t)]-
\Bigg[\sum_{l=L,R} \int dq\;\Big[\hat{\mathfrak{T}}^{l}(q),
\hat{\Omega}^{l}(q,t)\hat{\rho}_{S}(t) \nonumber \\ &&-
f(\epsilon^{l}(q)) \left\{ \hat{\rho}_{S}(t),\hat{\Omega}^{l}(q,t)
\right\} \Big] +\mathrm{H.c.}\Bigg]
\end{eqnarray}
with
\begin{eqnarray}
 \hat{\Omega}^{l}(q,t) &=& \frac{1}{\hbar^2}\chi^{l}(t)\exp\left(-\frac{i}{\hbar}t\epsilon^{l}(q)\right) \nonumber \\
 && \times \hat{U}_{S}(t) \hat{\Pi}^{l}(q,t)  \hat{U}_{S}^{\dagger}(t),
\end{eqnarray}
\begin{eqnarray}
 \hat{\Pi}^{l}(q,t)&=& \int_{0}^{t}dt'\; \left[ \exp\left(\frac{i}{\hbar}t'\epsilon^{l}(q)\right) \chi^{l}(t') \right. \nonumber \\
 && \times \left. \hat{U}_{S}^{\dagger}(t') \hat{\mathfrak{T}}^{l\dagger}(q) \hat{U}_{S}(t') \right]
\end{eqnarray}
and $f(E)$ being the Fermi distribution function.

Comparing this equation to the corresponding NZ equation,
\cite{Nakajima, Zwanzig, 1367-2630-11-11-113007, PhysRevB.81.155442}
\begin{eqnarray}
 \dot{\hat{\rho}}_{S}^{\rm NZ}(t)&=&-\frac{i}{\hbar}[\hat{H}_{S},\hat{\rho}_{S}^{\rm NZ}(t)]\nonumber \\
 &&-\Bigg[\sum_{l=L,R} \int dq\;[\hat{\mathfrak{T}}^{l}(q),\hat{\Omega}^{l}(q,t)]+\mathrm{H.c.}\Bigg] \label{NZ:RDO}
\end{eqnarray}
with
\begin{eqnarray}
 \hat{\Omega}^{l}(q,t)&=&  \frac{1}{\hbar^2}\chi^{l}(t)\hat{U}_{S}(t)  \int_{0}^{t}dt'\; \left[\exp\left(\frac{i}{\hbar}(t'-t)\epsilon^{l}(q)\right) \right. \nonumber \\ && \times \left. \chi^{l}(t')\hat{\Pi}^{l}(q,t')\right]\hat{U}_{S}^{\dagger}(t)
\end{eqnarray}
and
\begin{eqnarray}
 \hat{\Pi}^{l}(q,t')&=& \hat{U}_{S}^{\dagger}(t') \Big[\hat{\mathfrak{T}}^{l\dagger}(q)\hat{\rho}_S^{\rm NZ}(t')\nonumber \\&&-f(\epsilon^{l}(q))\left\{\hat{\rho}_S^{\rm NZ}(t'),\hat{\mathfrak{T}}^{l\dagger}(q)\right\}\Big]\hat{U}_{S}(t'), \label{NZ:Pi}
\end{eqnarray}
we note that we reobtain the TCL equation, if we set
\begin{equation}
 \hat{\rho}_S^{\rm NZ}(t') = \hat{U}_{S}^{\dagger}(t-t')\hat{\rho}_S(t)\hat{U}_{S}(t-t'),
\end{equation}
 in \eq{NZ:Pi} (which enters the kernel of \eq{NZ:RDO}), but let $\hat{\rho}_S^{\rm NZ}(t)=\hat{\rho}_S(t)$
in the first term of \eq{NZ:RDO}. In other words, in the Schr\"{o}dinger picture, the
NZ kernel takes the central system time propagated RDO (which lets
it become convoluted), while the TCL kernel takes just the
unpropagated RDO. The deviation between the two approaches is
therefore only of relevance when the central system is far from a
steady state and when the coupling to the leads is strong. It is our
experience that the positivity conditions~\cite{Whitney08:175304}
for the MB state occupation probabilities in the RDO are satisfied
to a higher system-lead coupling strength in the TCL case. The more
involved quantum structure demands a stronger system-lead coupling
than in our earlier work.~\cite{Gudmundsson85:075306} The numerical
effort of the two approaches is of similar magnitude. Both cases
allow for a $t$-independent inner time integral over $t'$, which can
be integrated successively with increasing $t$ (increasing
integration domain).~\cite{1367-2630-11-7-073019} The RDO is inside
(NZ) or outside (TCL) of the inner time integral, but the required
number of matrix multiplications is equal.

\section{Non-equilibrium transport properties}

In this section, we investigate the non-equilibrium electron
transport properties through a quantum ring system, 
which is situated in a photon
cavity and weakly coupled to leads. We assume GaAs-based
material with electron effective mass $m^*=0.067m_e$ and background
relative dielectric constant $\kappa = 12.4$. We consider a single cavity mode
with fixed photon excitation energy $\hbar\omega = 0.4$~meV. The
electron-photon coupling constant in the central system is $g^{EM} =
0.1$~meV.  Before switching on the coupling, we assume the central
system to be in the pure initial state with electron occupation
number $N_{e,\rm{init}}=0$ and photon occupation
number $N_{ph,\rm{init}}=1$ of the electromagnetic field.

An external perpendicular uniform magnetic field is applied through
the central ring system and the lead reservoirs.  The area of the
central ring system is $A \approx 2\times 10^{4}$~$\mathrm{nm}^2$ so
that the magnetic field corresponding to the flux quantum
$\Phi_0$ is $B_0=\Phi_0/A \approx 0.2$~T. The temperature of
the reservoirs is assumed to be T = 0.5 K. The chemical potentials
in the leads are $\mu_L=2$~meV and $\mu_R=0.9$~meV leading to a
source-drain bias window $\Delta \mu = 1.1$~meV.  To facilitate
inelastic scattering processes between the SES in the central
ring system and the SES in the lead $l$, we allow for coupling of
highly energetically different states by letting the affinity
constant $\Delta_E^{l} = 4.0$~meV.\cite{Abdullah10:195325}   In
addition, we let the contact region parameters for lead $l\in\{L,R\}$ in x-
and y-direction be $\delta^l_x = \delta^l_y= 4.39\times
10^{-4}$~$\rm{nm}^{-2}$.  The system-lead coupling strength $g_0^l =
0.2058$~${\rm meV}/{\rm nm}^{3/2}$.

There are several relevant length and time scales that should be
mentioned.  The two-dimensional magnetic length is $l =
[c\hbar/(eB)]^{1/2} = 25.67 [B({\rm T})]^{-1/2}$~nm.  The ring system
is parabolically confined in the y-direction with characteristic
energy $\hbar\Omega_0 = 1.0$~meV leading to a modified magnetic
length scale
\begin{align}
a_w &= \left(\frac{\hbar}{m^*\Omega_0}\right)^{1/2}
    \frac{1}{\sqrt[4]{1+[eB/(m^*c\Omega_0)]^2}}\nonumber\\
    &= \frac{33.74}{ \sqrt[4]{1+2.982[B({\rm T})]^2} }\ {\rm nm}.
\end{align}
Correspondingly, the system-lead coupling strength is then $g_0^l
a_w^{3/2} = 39.85$~meV for magnetic field $B=0.1$~T and $g_0^l
a_w^{3/2} = 38.22$~meV for magnetic field $B=0.225$~T. The
time-scale for the switching on of the system-lead coupling is
$(\alpha^{l})^{-1}=3.291$~\textrm{ps}, the single-electron state
(1ES) charging time-scale $\tau_{\rm 1ES}\approx 30$~\textrm{ps},
and the two-electron state (2ES) charging time-scale $\tau_{\rm 2ES}
\gg 200$~\textrm{ps} described in the sequential tunneling regime. 
We study the transport properties for $0\leq t<\tau_{\rm 2ES}$,
when the system has not yet reached a steady state.

In order to understand the non-equilibrium dynamical behavior of the
charge distribution in the system, we define the time-dependent magnitude of
charge on the left part ($x<0$) of the ring
\begin{equation}
 Q_S^{L}(t)=\int_{-\frac{L_x}{2}}^{0}dx\;\int_{-\infty}^{\infty}dy\;\rho(\mathbf{r},t)
\end{equation}
and the time-dependent magnitude of charge on the right part ($x>0$)
of the ring
\begin{equation}
 Q_S^{R}(t)= \int_{0}^{\frac{L_x}{2}}dx\;\int_{-\infty}^{\infty}dy\;\rho(\mathbf{r},t)\,
 .
\end{equation}
The space- and time-dependent charge density,
\begin{equation}
 \rho(\mathbf{r},t)=\mathrm{Tr}[\hat{\rho}_{S}(t)\hat{\rho}(\mathbf{r})],
\end{equation}
is the expectation value of the charge density operator
\begin{equation}
 \hat{\rho}(\mathbf{r})=e\hat{\psi}^{\dagger}(\mathbf{r})
 \hat{\psi}(\mathbf{r}).
\end{equation}

In order to explore the magnetic field influence on the charge
currents from and into the leads, 
we define the charge current from the left lead into the
system by
\begin{equation}
 I_{L}(t)=\mathrm{Tr}[\dot{\hat{\rho}}_S^L(t) \hat{Q}]\, .
\end{equation}
Here,  $\hat{Q}=e\hat{N}$ is the charge operator with number
operator $\hat{N}$ and the time-derivative of the RDO in the MB
basis due to the coupling to the lead $l\in\{L,R\}$
\begin{eqnarray}
 \dot{\rho}_S^l(t)&=&\int dq\;\Big[\mathfrak{T}^{l}(q), \Big[
\Omega^{l}(q,t)\rho_{S}(t) - \nonumber \\
 && f(\epsilon^{l}(q)) \left\{ \rho_{S}(t),\Omega^{l}(q,t) \right\}
\Big] \Big]+\mathrm{H.c.}.
\end{eqnarray}
Similarly, the charge current from the system into the right lead
can be expressed as
\begin{equation}
 I_{R}(t)=-\mathrm{Tr}[\dot{\hat{\rho}}_S^R(t) \hat{N}].
\end{equation}
To get more insight into the local current flow in the ring system,
we define the top local charge current through the upper arm ($y>0$)
of the ring
\begin{equation}
 I_{\mathrm{top}}(t)=\int_{0}^{\infty}dy\;j_x(x=0,y,t)
\end{equation}
and the bottom local charge current through the lower arm ($y<0$) of
the ring
\begin{equation}
 I_{\mathrm{bottom}}(t)=\int_{-\infty}^{0}dy\;j_x(x=0,y,t)\, .
\end{equation}
Here, the charge current density,
\begin{equation}
 \mathbf{j}(\mathbf{r},t)
 =\begin{pmatrix} j_x(\mathbf{r},t)\\j_y(\mathbf{r},t) \end{pmatrix}
 =\mathrm{Tr}[\hat{\rho}_{S}(t)\hat{\mathbf{j}}(\mathbf{r})],
\end{equation}
is given by the expectation value of the charge current density
operator,
\begin{equation}
 \hat{\mathbf{j}}(\mathbf{r})=\hat{\mathbf{j}}_{p}(\mathbf{r})+\hat{\mathbf{j}}_{d}(\mathbf{r}),
\end{equation}
decomposed into the paramagnetic charge current density operator,
\begin{equation}
 \hat{\mathbf{j}}_{p}(\mathbf{r})=\frac{e\hbar}{2mi}\left[\hat{\psi}^{\dagger}(\mathbf{r})(\nabla \hat{\psi}(\mathbf{r}))-
 (\nabla \hat{\psi}^{\dagger}(\mathbf{r})) \hat{\psi}(\mathbf{r}) \right],
\end{equation}
and the diamagnetic charge current density operator,
\begin{equation}
 \hat{\mathbf{j}}_{d}(\mathbf{r})=\hat{\mathbf{j}}_{d}^{\rm mag}(\mathbf{r})+\hat{\mathbf{j}}_{d}^{\rm ph}(\mathbf{r}).
\end{equation}
The latter consists of a magnetic component,
\begin{equation}
 \hat{\mathbf{j}}_{d}^{\rm mag}(\mathbf{r})=\frac{e^2}{m}\mathbf{A}(\mathbf{r})\hat{\psi}^{\dagger}(\mathbf{r})\hat{\psi}(\mathbf{r}),
\end{equation}
and photonic component,
\begin{equation}
 \hat{\mathbf{j}}_{d}^{\rm ph}(\mathbf{r})= \frac{e^2}{m}\hat{\mathbf{A}}^{\mathrm{ph}}(\mathbf{r})\hat{\psi}^{\dagger}(\mathbf{r})\hat{\psi}(\mathbf{r}).
\end{equation}
Furthermore, to understand better the driving schemes of the dynamical transport features,
 we define the total local charge current
\begin{equation}
 I_{\rm tl}(t)=I_{\rm top}(t)+I_{\rm bottom}(t)
\end{equation}
and circular local charge current
\begin{equation}
 I_{\rm cl}(t)=\frac{1}{2}(I_{\rm top}(t)-I_{\rm bottom}(t)).
\end{equation}
Below, we shall explore the influence of the applied magnetic field and the
photon fiel polarization on the non-equilibrium
quantum transport in terms of the above time-dependent charges and
currents in the broad quantum ring system connected to leads.

\subsection{Photons with x-polarization}

\begin{figure}[htbq] 
       \includegraphics[width=0.45\textwidth,angle=-90]{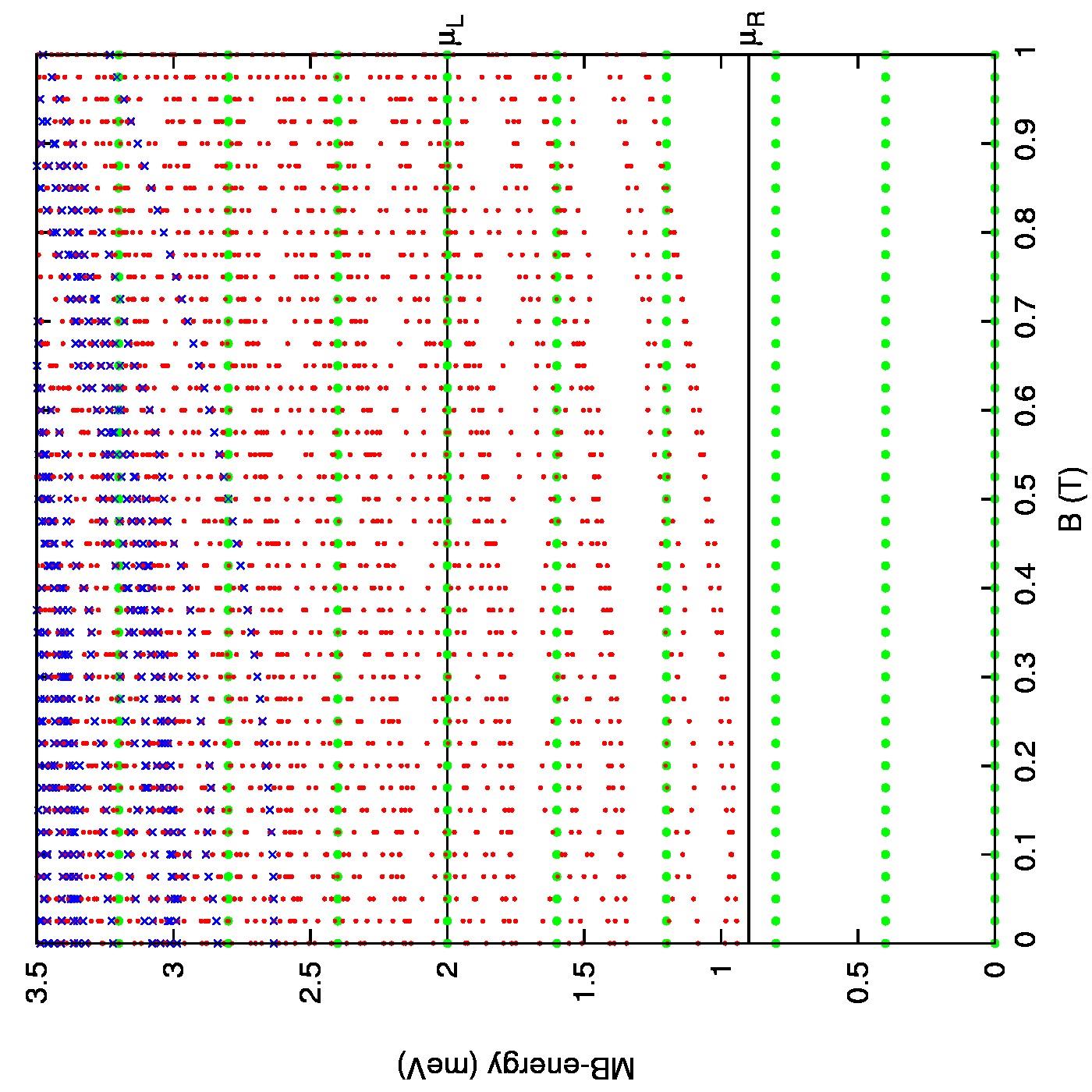}
       \caption{(Color online)
       MB energy spectrum of system Hamiltonian $\hat{H}_{S}$ versus magnetic field $B$
       in units of tesla (T).
       The states are differentiated according to their electron content $N_e$:
       zero-electron states ($N_e=0$, 0ES, green dots), single electron states ($N_e=1$, 1ES, red dots)
       and two electron states ($N_e=2$, 2ES, blue crosses). The photon field is x-polarized. }
       \label{MB_spec_vB_x}
\end{figure}

In this subsection, we focus on our results for {\em x}-polarized photon field.
Figure \ref{MB_spec_vB_x} shows the MB energy spectrum of the system
Hamiltonian $\hat{H}_{S}$ including the electron-electron and
electron-photon interactions. The MB-energy levels are assigned
different colors according to their electron content $N_e$. For
$N_e=0$ (green dots), the MB states differ in energy by multiples of the photon energy
$\hbar \omega$ according to their photon content $N_{ph}$ 
independently of the applied magnetic field. The bias window contains a number of SES
(red dots) of which the two lowest ones have the highest occupation.
The SESs show crossing behavior at half integer flux quanta. The
state-dependent effective magnetic flux ring area allows for small variations of the
crossing period $B_0$. The crossings at integer flux quanta
are usually avoided due to the ring rotation symmetry violation, 
which is mainly coming from the presence of the central system contact regions to the leads.
In general, the SES and, in particular, the two
electron states (2ES, blue crosses) tend to increase in energy with
higher magnetic field.

\begin{figure}[htbq] 
       \includegraphics[width=0.34\textwidth,angle=-90]{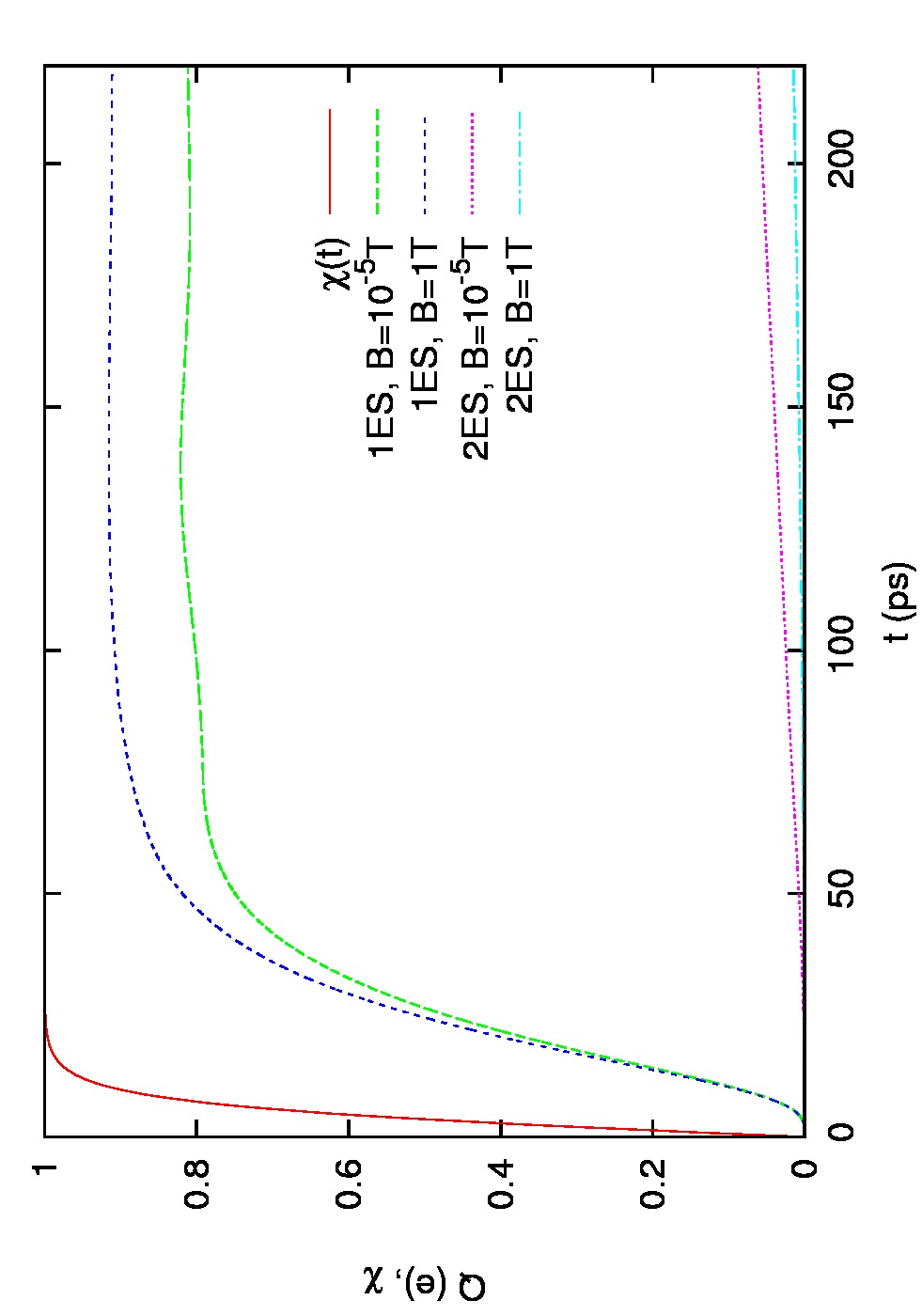}
       \caption{(Color online) Switching function $\chi^{l}(t)$ (solid red),
       charge of all 1ES for $B=10^{-5}$~T (dashed green) and
       $B=1.0$~T (dotted blue), and charge of all 2ES for $B=10^{-5}$~T (dotted purple) and
       $B=1.0$~T (dash-dotted cyan)
       as a function of time. The photon field is x-polarized. }
       \label{1ES_2ES_x}
\end{figure}

Figure \ref{1ES_2ES_x} illustrates the central region charging of 1ES and 2ES as a
function of time with the initial conditions $Q_{\rm 1ES}(0)=Q_{\rm 2ES}(0)=0$
since we selected an initial state with $N_{e,\rm{init}}=0$.  
In the low magnetic field regime, in the case
$B=10^{-5}$~T, we notice that $(t,Q_{\rm 1ES}) = (200\, {\rm ps},
0.809e)$ and $(t,Q_{\rm 2ES}) = (200\, {\rm ps}, 0.055e)$. In the
high magnetic field regime, in the case $B=1$~T, we notice that
$(t,Q_{\rm 1ES}) = (200\, {\rm ps}, 0.911e)$ and $(t,Q_{\rm 2ES}) =
(200\, {\rm ps}, 0.012e)$. In general, the 2ES are occupied slower
than the 1ES indicating the sequential tunneling processes. 
But more importantly, the energetic location of the 2ES and their shifting above
the bias window by the Coulomb interaction attenuates the 2ES occupation.
As is shown in \fig{MB_spec_vB_x}, the magnetic
field plays a role to increase further the energy difference of the 2ES with
respect to the 1ES, thus enhancing the 1ES occupation by $\delta Q_{\rm 1ES} =
0.102e$ while reducing the 2ES occupation by $\delta Q_{\rm 2ES} =
-0.043e$. We note that the earlier mentioned time-scales $(\alpha^{l})^{-1}=3.291$~\textrm{ps},
$\tau_{\rm 1ES}\approx 30$~\textrm{ps} and $\tau_{\rm 2ES}
\gg 200$~\textrm{ps} are in agreement with \fig{1ES_2ES_x}.

\begin{figure}[htbq]
       \includegraphics[width=0.34\textwidth,angle=-90]{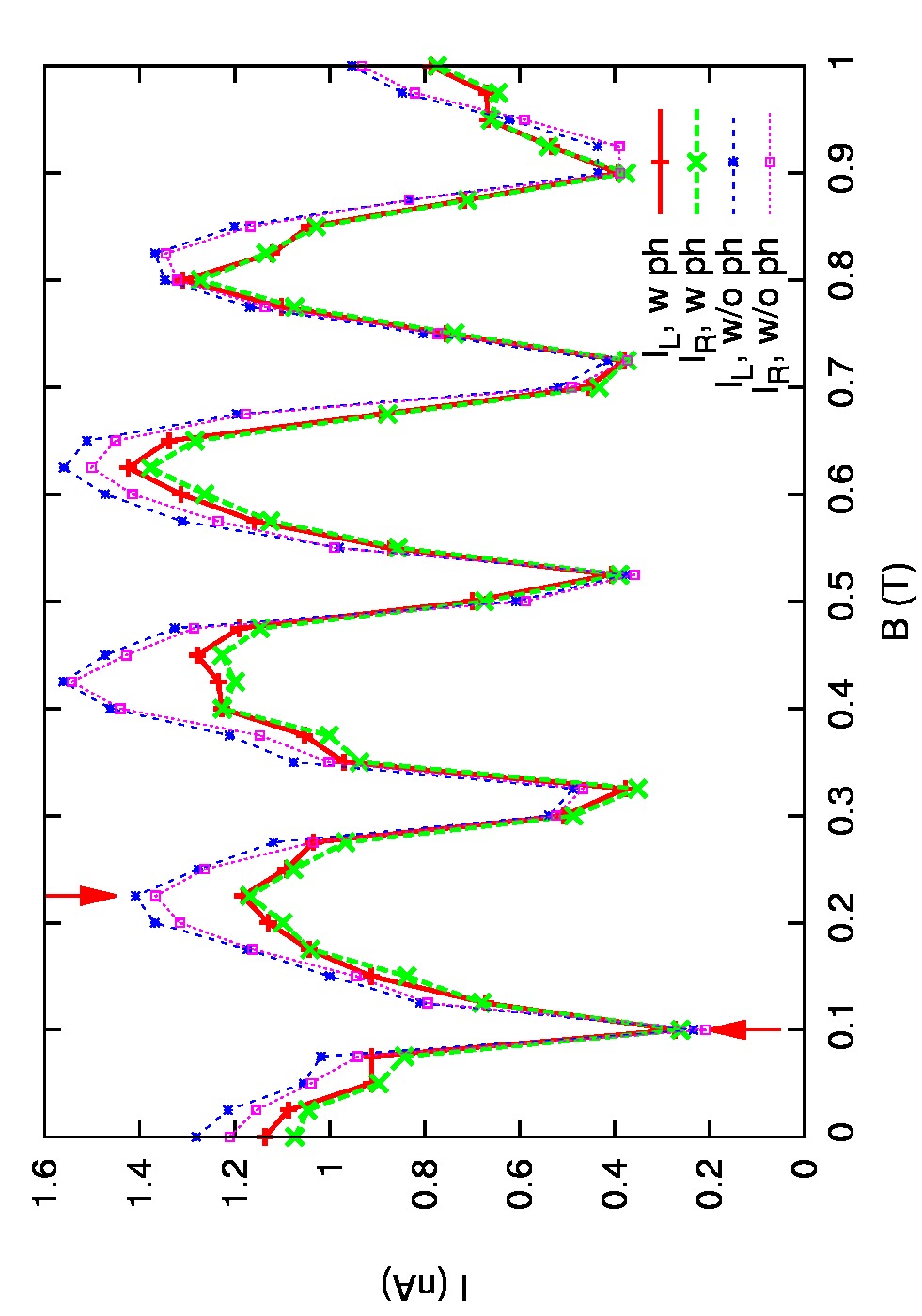}
       \caption{(Color online) The left charge current $I_{L}$ (solid red) and the right charge current
       $I_R$ (long-dashed green) versus the magnetic field with (w) x-polarized photon field
       at $t=200\mathrm{ps}$. For comparison: left charge current $I_{L}$ (short-dashed blue) and right charge current
       $I_R$ (dotted purple) in a purely electronic central system, i.e. without (w/o) photon cavity.}
       \label{current_per_200_x}
\end{figure}

In \fig{current_per_200_x}, we show the current from the left lead
into the ring system  $I_{L}$ (solid red curve) and the current from the ring system
to the right lead $I_R$ (long-dashed green curve) as a function of magnetic field at time
$t=200$~ps. The similar values of $I_L(B)$ and $I_R(B)$ indicate
that the short-time regime charging of 1ES is nearly completed
at $t=200$~ps meaning that the total charging has slowed down by more than an order of magnitude.
Moreover, we see clear oscillations of the current with period
$B_0\approx 0.2$~T: the first minimum current at $B=0.1$~T
corresponds to the situation of a half flux quantum, in which the
left charge current $I_L = 0.273$~nA and the right charge current
$I_R = 0.261$~nA and the maximum current at $B=0.225$~T is
corresponding to the case of one flux quantum, in which the left
charge current $I_L=1.183$~nA and the right charge current
$I_R=1.168$~nA. These observations are in agreement with the
Aharonov-Bohm (AB) oscillations of the steady
state~\cite{PhysRev.115.485,Buttiker30.1982,Webb.54.2696} but
superimposed and modified by electron-electron correlation effects
and the non-equilibrium situation. In addition, the electron-photon
coupling suppresses the constructive interference of AB-phases in
the integer flux quantum situation as can be seen from a comparison
with the purely electronic system results in \fig{current_per_200_x}
(short-dashed blue and dotted purple curve).

\begin{figure}[htbq]
       \subfigure[]{
       \includegraphics[width=0.34\textwidth,angle=-90]{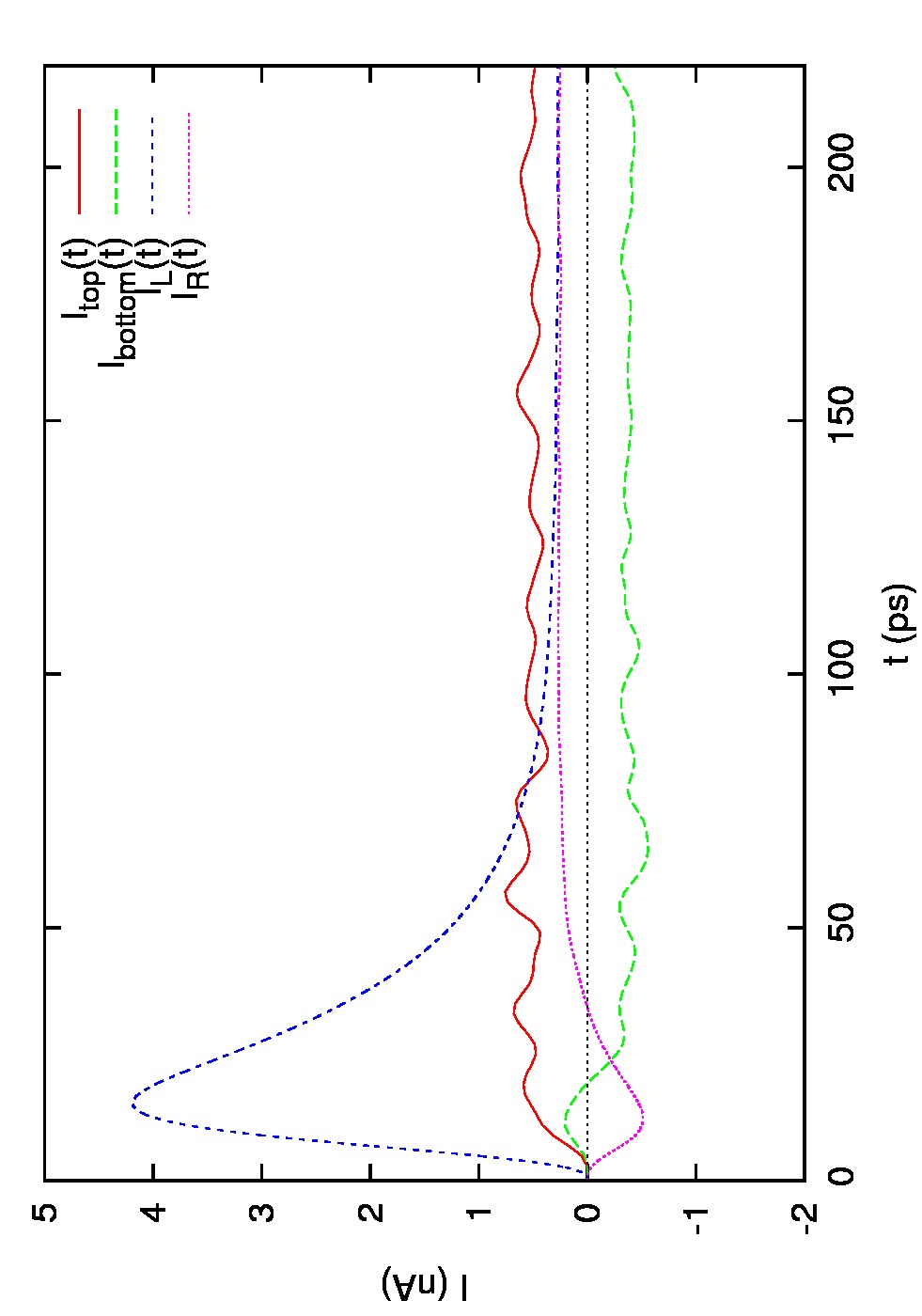}
       \label{current_all_x_vt_B010}}
       \subfigure[]{
       \includegraphics[width=0.34\textwidth,angle=-90]{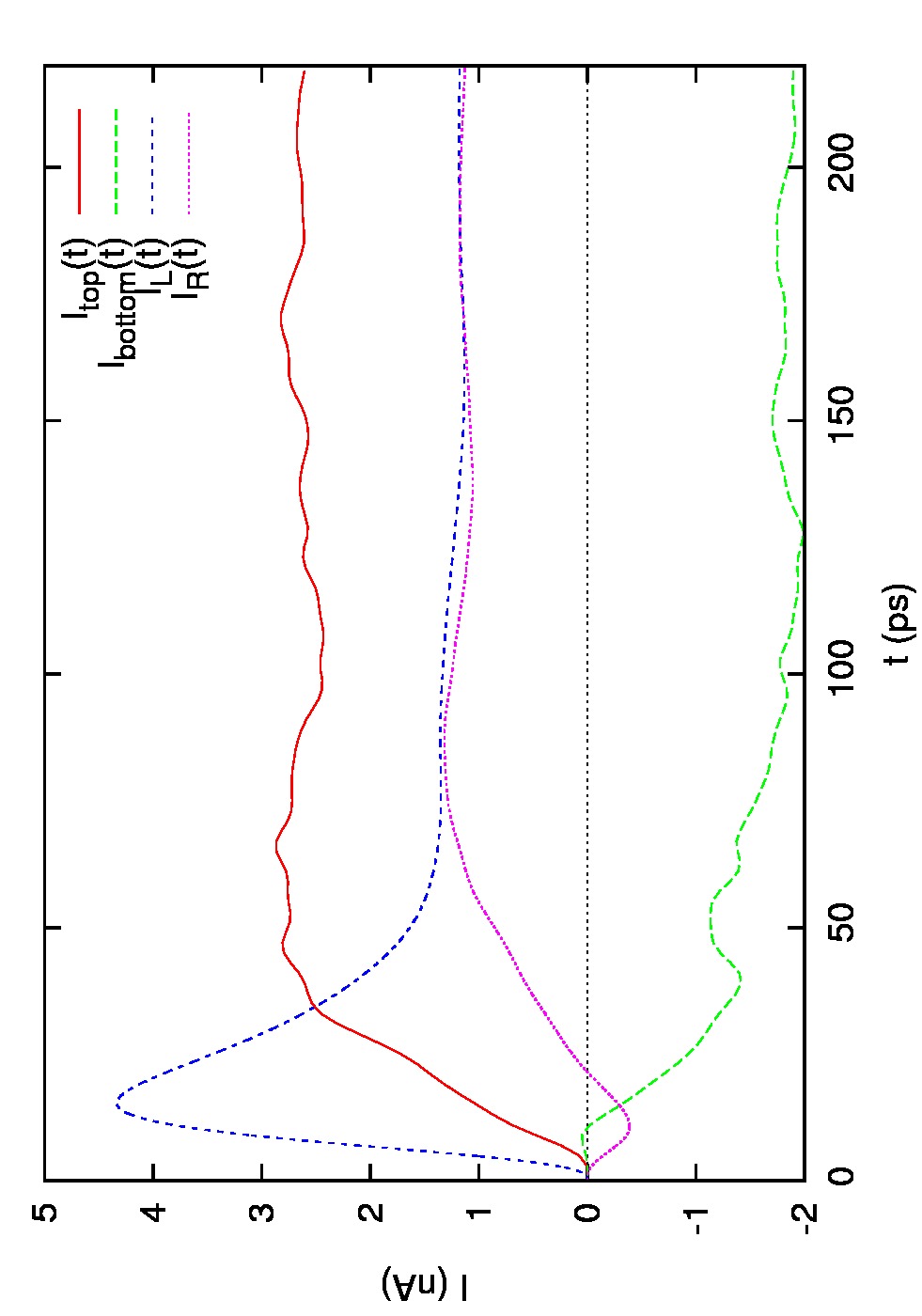}
       \label{current_all_x_vt_B0225}}
       \caption{(Color online) Local current through the top ring branch
       $I_{\mathrm{top}}(t)$,
       local current through the bottom ring branch $I_{\mathrm{bottom}}(t)$,
       current from the left lead into the system $I_{L}(t)$,
       and current from the system into the right lead $I_R(t)$
       as a function of time for \subref{current_all_x_vt_B010}
       $B=0.1$~T and \subref{current_all_x_vt_B0225}
       $B=0.225$~T in the case of x-polarized photon field.}
       \label{current_all_x_vt}
\end{figure}

Figure \ref{current_all_x_vt} shows the time-evolution of the left
total current $I_L(t)$, the right total current $I_R(t)$, the top
local current $I_{\mathrm{top}}(t)$, and the bottom local current
$I_{\mathrm{bottom}}(t)$. In the case of $B=0.1$~T (a half flux
quantum) shown in \fig{current_all_x_vt}(a), the maximum value of
the charge current from the left lead at $t=15.30$~ps is $I_L=4.190$~nA.
In addition, the minimum value of the charge current to the right lead at
$t=12.25$~ps is $I_R = -0.511$~nA. The negative right charge current
indicates that the central system is also charged from the right lead
during the transient phase. In the case of
$B=0.225$~T (one flux quantum) shown in \fig{current_all_x_vt}(b),
the maximum value of the current from the left lead at $t=15.35$~ps
is $I_L=4.334$~nA; and the minimum right charge current at
$t=10.80$~ps is $I_R=-0.387$~nA. It is then interesting to
realize that the magnetic field enhances the charge
accumulation from the left lead, while it suppresses the short-time
regime charging from the right lead. Hence, the integer magnetic
flux quantum case assists the net current flow 
already in the highly non-equilibrium situation in the very beginning. 
The local charge transport may differ in direction in the ring arms 
due to the persistent magnetic field induced ring current.

\begin{figure}[htbq]
       \includegraphics[width=0.37\textwidth,angle=0,viewport=40 8 218 170,clip]{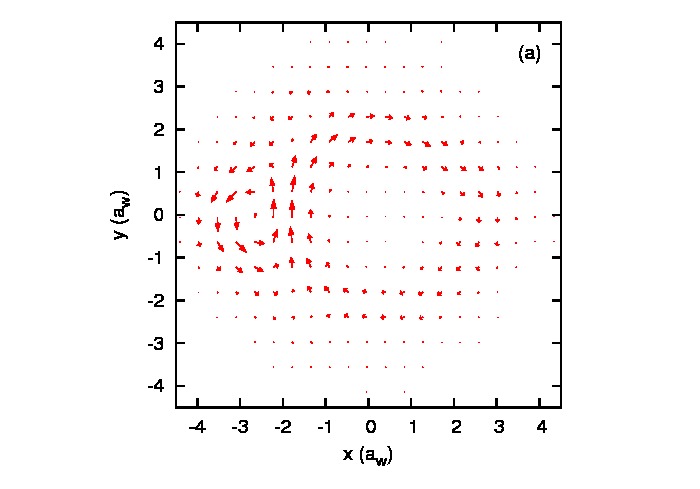}
       \includegraphics[width=0.37\textwidth,angle=0,viewport=40 8 218 170,clip]{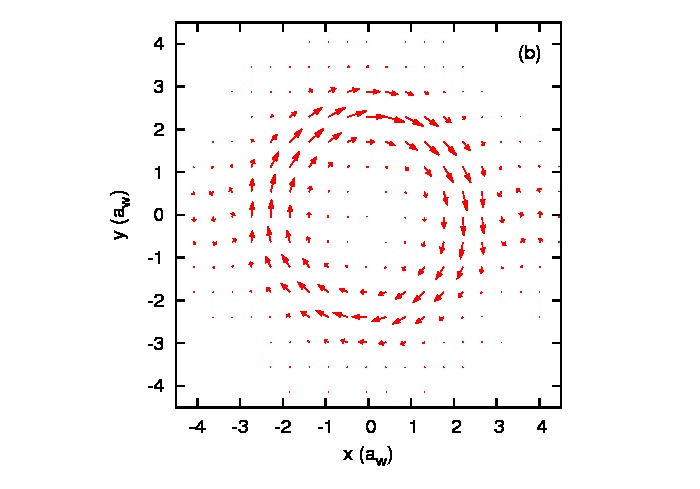}
       \includegraphics[width=0.37\textwidth,angle=0,viewport=40 8 218 170,clip]{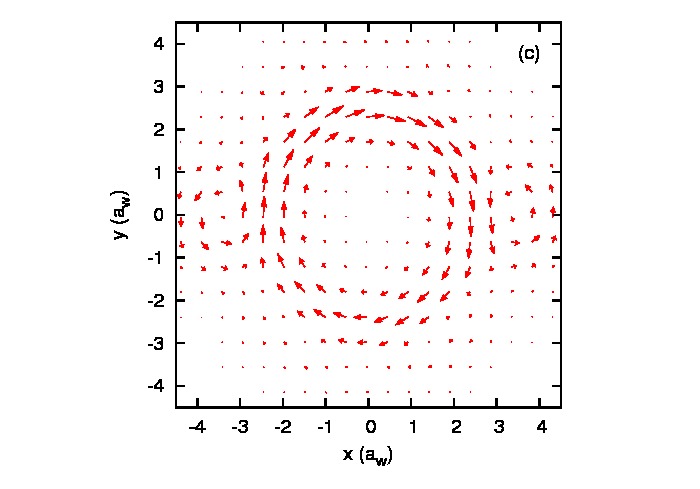}
       \caption{(Color online) Normalized charge current density vector field in the central system
       for (a) $B=0.1$~T, (b) $B=0.225$~T, and (c) $B=0.425$~T at $t=200$~ps
       in the case of x-polarized photon field.}
       \label{cur_den_x_200}
\end{figure}

In \fig{cur_den_x_200}, we illustrate the normalized charge current
density vector field $\mathbf{j}(\mathbf{r},t)$ in the central
quantum ring system in the case
of x-polarized photon field in the long-time response regime $t=200$~ps, i.e. when the 2ES get charged.
For magnetic field $B=0.1$~T, a clear
counter-clockwise vortex located close to the left
lead can be found dominating the current flow pattern in the
central ring system as shown in \fig{cur_den_x_200}(a).~\cite{Pichugin56:9662}
The vortex circulation direction is in agreement
with the Lorentz force, since the enclosed area is threaded by much
less than half a flux quantum. Due to the geometrical position of
the vortex and the current continuity condition, clockwise current
direction is favored for the ring system. However, the
counter-clockwise vortex appears relatively weak for magnetic field
$B=0.225$~T present at both left and right lead connection area as
shown in \fig{cur_den_x_200}(b), while the total local current
through the whole central system from the left to the right lead is
large. Additionally, for a later comparison with the y-polarized photon field,
\fig{cur_den_x_200}(c) shows the current density for $B=0.425$~T
(two flux quanta), which is similar to \fig{cur_den_x_200}(b) (one
flux quanta) with the vortex circulation on both sides being slightly
more significant.

\begin{figure}[htbq]
       \subfigure[]{
       \includegraphics[width=0.34\textwidth,angle=-90]{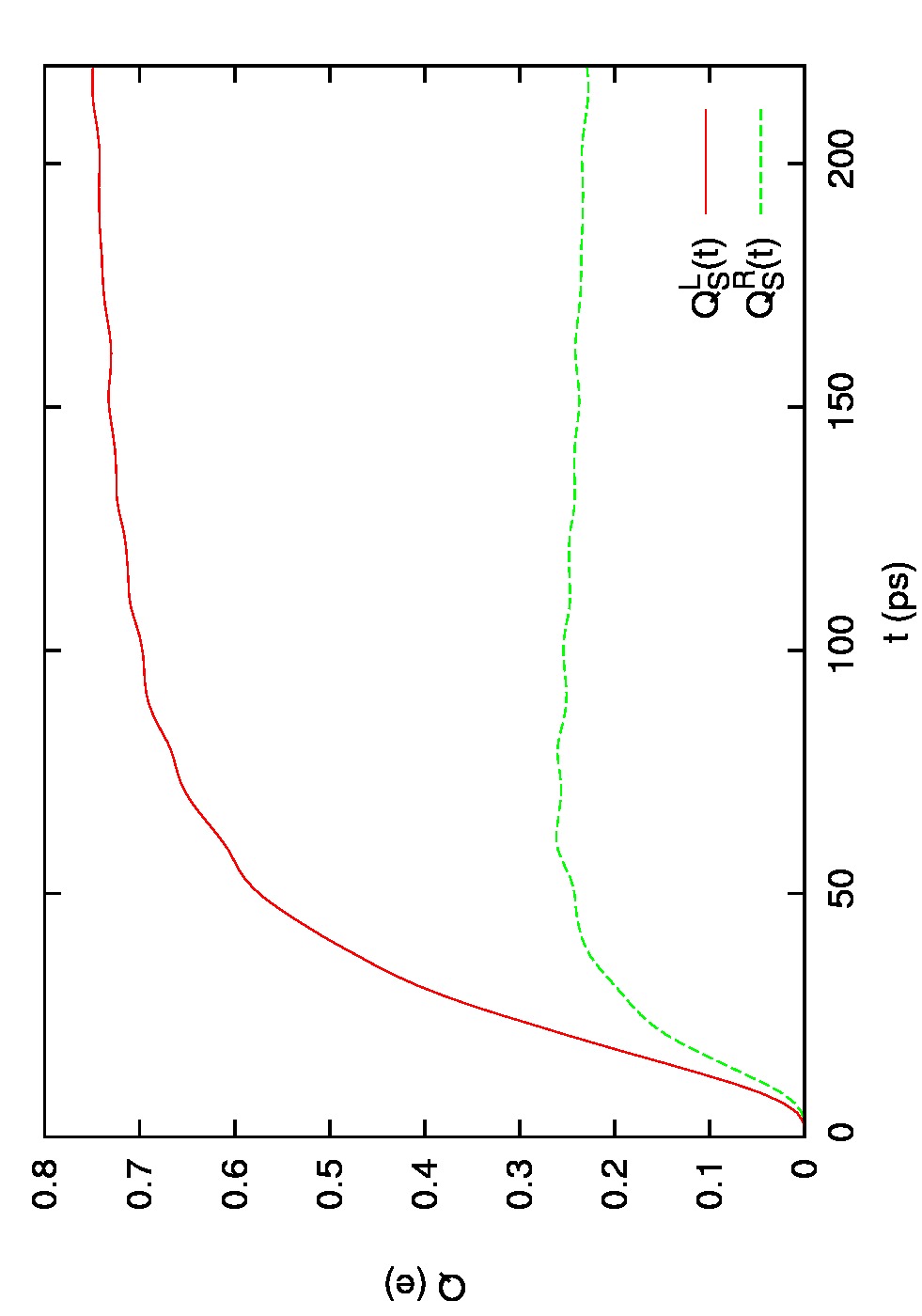}
       \label{density_all_x_vt_B010}}
       \subfigure[]{
       \includegraphics[width=0.34\textwidth,angle=-90]{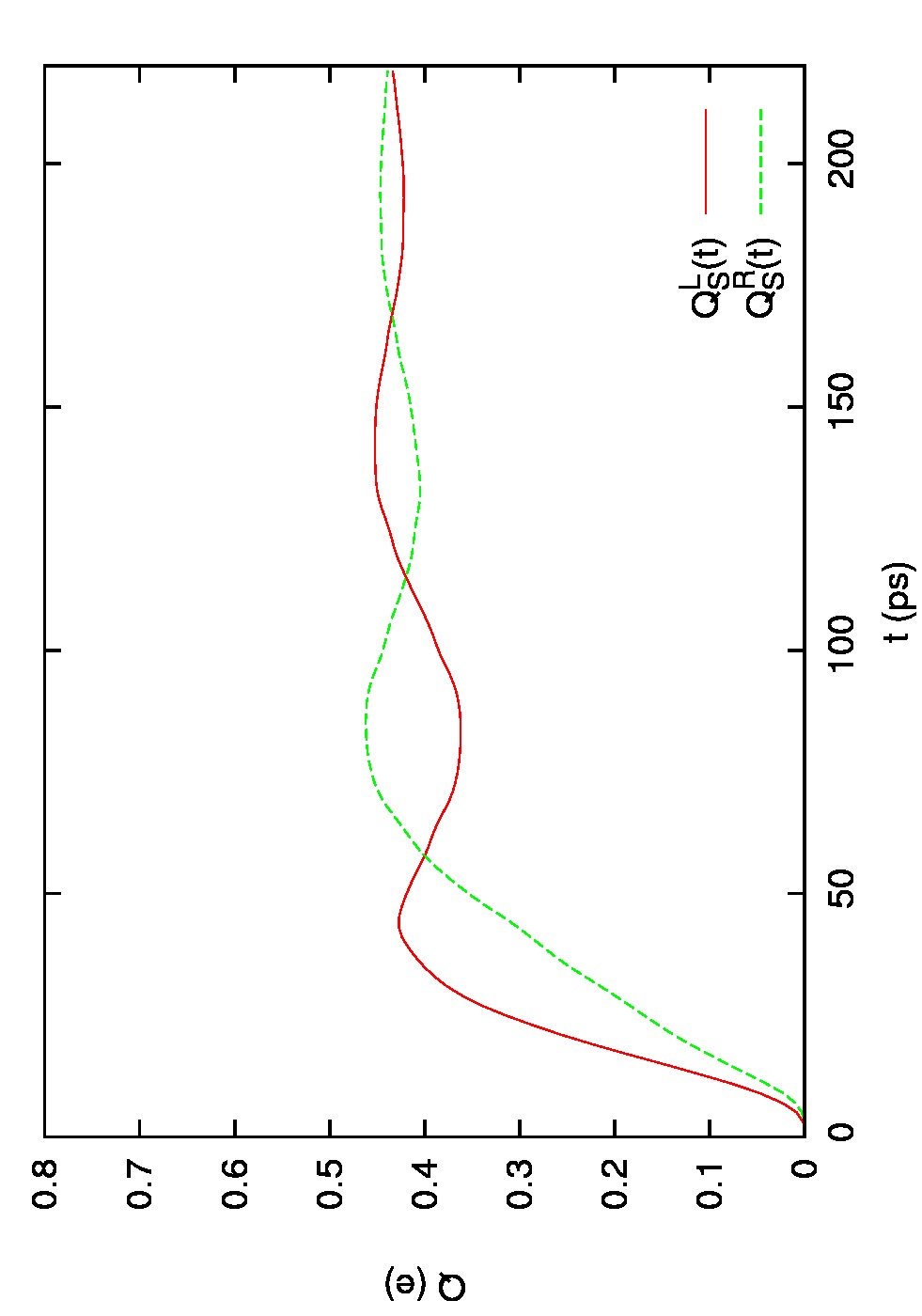}
       \label{density_all_x_vt_B0225}}
       \caption{(Color online) Charge in the left ($Q_S^{L}(t)$) or right half ($Q_S^{R}(t)$)
       of the central quantum ring system as a function of time
       for \subref{density_all_x_vt_B010} $B=0.1$~T and \subref{density_all_x_vt_B0225}
       $B=0.225$~T. The photon field is x-polarized.}
       \label{density_all_x_vt}
\end{figure}

Figure \ref{density_all_x_vt} illustrates the time-dependent charge
on the left part of the ring $Q_S^{L}(t)$ and the time-dependent
charge on the right part of the ring $Q_S^{R}(t)$. In the case of
$B=0.1$~T shown in \fig{density_all_x_vt}(a),  both $Q_S^{L}$ and
$Q_S^{R}$ are increasing almost monotonically in time.  In the
long-time response regime $t=200$~ps, $Q_S^{L}(t)=0.742e$ is much
higher than $Q_S^{R}(t)=0.234e$. This implies charge accumulation mainly on
the left hand side of the quantum ring in the case of magnetic field
with half integer flux quantum with enhancement of the
electron dwell time on the left-hand side of the ring and suppression of the
electron dwell time on the right-hand side. 

In the case $B=0.225$~T shown in
\fig{density_all_x_vt}(b), both $Q_S^{L}$ and $Q_S^{R}$ exhibit
oscillations in time after the short-time charging regime. This
implies that the charge accumulation manifests itself in oscillating
behavior between the left and the right part of the quantum ring in
the case of magnetic field with integer flux quantum. 
The oscillation amplitude is decreasing in
time due to the dissipation effects caused by the coupling to the
leads. In the long-time response regime $t=200$~ps, $Q_S^{L}(t)=0.423e$ is
of similar magnitude than $Q_S^{R}(t)=0.446e$, which is by difference
to the half integer flux quantum case.
It is interesting
to notice that the oscillating period of the charges is around
$\tau=100$~ps corresponding to a characteristic energy $\delta E_Q
\approx 0.04$~meV. The MB energies of the mostly occupied MB
levels are $E_{10}^{x}=1.4038$~meV $E_{9}^{x}=1.3664$~meV such that $\Delta
E_{9,10}^{x}=0.0374$~meV. The corresponding two-level (TL)
oscillation period of the closed system would be $\tau_{\rm
TL}^{0}=111$~ps. In the non-equilibrium open system, the TL
oscillation period is $\tau_{\rm TL}^{L}=94$~ps or $\tau_{\rm
TL}^{R}=100$~ps when we take the time intervals between the first
and second maxima of $Q_S^{L}(t)$ and $Q_S^{R}(t)$, respectively.
The full numerical calculation including all MB levels shown in \fig{density_all_x_vt}(b)
yields the left and right charge oscillation period, $\tau^L=96$~ps and
$\tau^R=110$~ps, respectively. The system is far from equilibrium at
the earlier maximum, thus reducing in particular the left period
$\tau_{\rm TL}^L$ with respect to $\tau_{\rm TL}^{0}$. However, we
find that also the other MB states change the periods when comparing
$\tau^L$ with $\tau_{\rm TL}^{L}$ and $\tau^R$ with $\tau_{\rm
TL}^{R}$.

We would like to bring attention to the fact that
charge balances like $\dot{Q}_{L}= I_{L}-I_{\rm tl}$ and $
\dot{Q}_{R} = I_{\rm tl} - I_{R}$ would not be satisfied. This is because
the SES that are filled from the left lead or emptied to the right
lead are in general not restricted to a single half of the central
system, but extended over the whole system.

\begin{figure}[htbq]
       \subfigure[]{
       \includegraphics[width=0.37\textwidth,angle=0,viewport=45 12 240 152,clip]{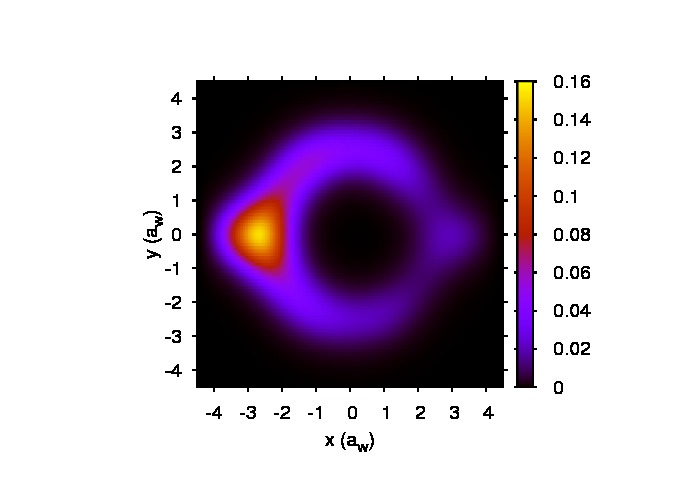}
       \label{den_B010x_200}}
       \subfigure[]{
       \includegraphics[width=0.37\textwidth,angle=0,viewport=45 12 240 152,clip]{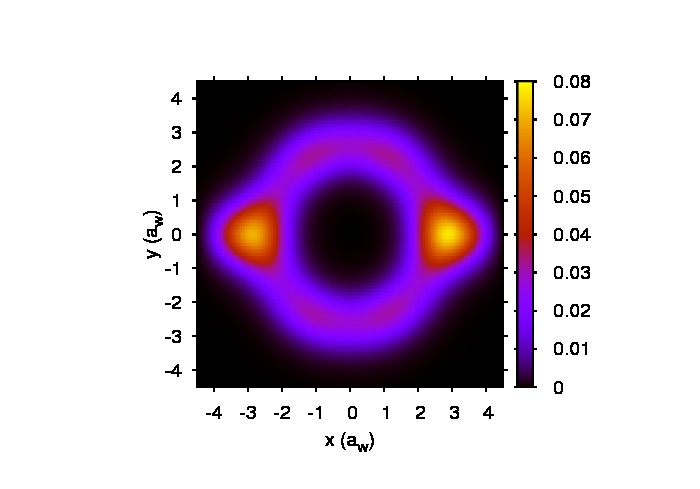}
       \label{den_B0225x_200}}
       \subfigure[]{
       \includegraphics[width=0.37\textwidth,angle=0,viewport=45 12 240 152,clip]{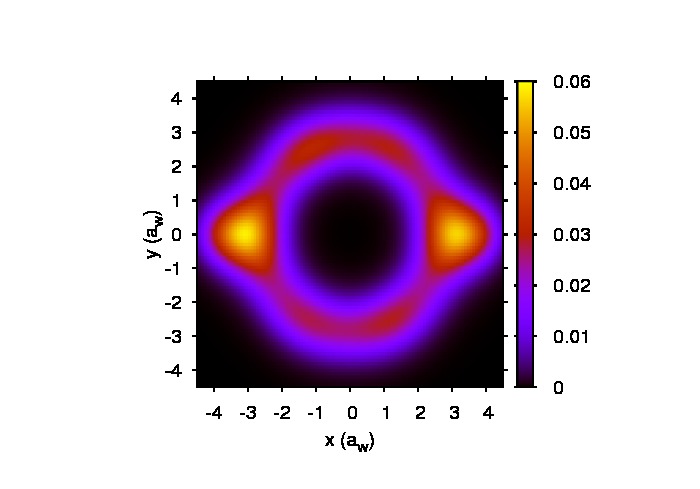}
       \label{den_B0425x_200}}
       \caption{(Color online) Charge density distribution $\rho(\mathbf{r},t)\; (e/a_w^2)$ in the central system
       for \subref{den_B010x_200} $B=0.1$~T, \subref{den_B0225x_200} $B=0.225$~T, and
       \subref{den_B0425x_200} $B=0.425$~T in the x-polarized photon field case at $t=200$~ps.}
       \label{den_x_200}
\end{figure}

Figure \ref{den_x_200} shows the charge density distribution in the
central quantum ring system in the case of x-polarized photon field
with the magnetic field \subref{den_B010x_200} $B=0.1$~T,
\subref{den_B0225x_200} $B=0.225$~T, and \subref{den_B0425x_200}
$B=0.425$~T at $t=200$~ps. In the case of $B=0.1$~T (half flux
quantum) shown in \fig{density_all_x_vt}\subref{den_B010x_200}, the
electrons are highly accumulated on the left-hand side of the
quantum ring with very weak coupling to the right lead, and hence
strongly blocking the left charge current and suppressing the right
charge current, as it was shown previously in \fig{current_per_200_x}
(marked by the up-arrow).  For half integer flux quantum, the
electron dwell time on the left-hand side of the ring is enhanced
relative to the electron dwell time on the right-hand side of the
ring due to destructive phase interference on the right hand side.
The density accumulates then mainly on the left hand side of the
ring forming a long-living state and the magnetic field evoked
vortex on the right hand side contact area is suppressed.

In the $B=0.225$~T case (one flux quantum) shown in
\fig{den_x_200}\subref{den_B0225x_200}, the electrons manifest
oscillating feature between the left and right end of the quantum
ring. At time $t=200$~ps, the electrons are nearly equally well accumulated
on both sides of the quantum ring.  This phenomenon is related to
the manifestation of current peaks observed in
\fig{current_per_200_x} (marked by the down-arrow). The charge
distribution is rearranged when compared to magnetic field $B=0.1$~T. We
observe a depletion of about $50\%$ at the left-hand contact region with
equivalent charge augmentation on the right-hand contact region. The magnetic
field $B=0.225$~T with integer flux quantum enhances the likelihood
for electrons to flow through the quantum ring to the right-hand
side of the central system and further to the right lead.
Additionally, \fig{den_x_200}\subref{den_B0425x_200} shows the
charge density for $B=0.425$~T (two flux quanta), which is similar
to \fig{den_x_200}\subref{den_B0225x_200} (one flux quantum).

\begin{figure}[htbq]
       \includegraphics[width=0.34\textwidth,angle=-90]{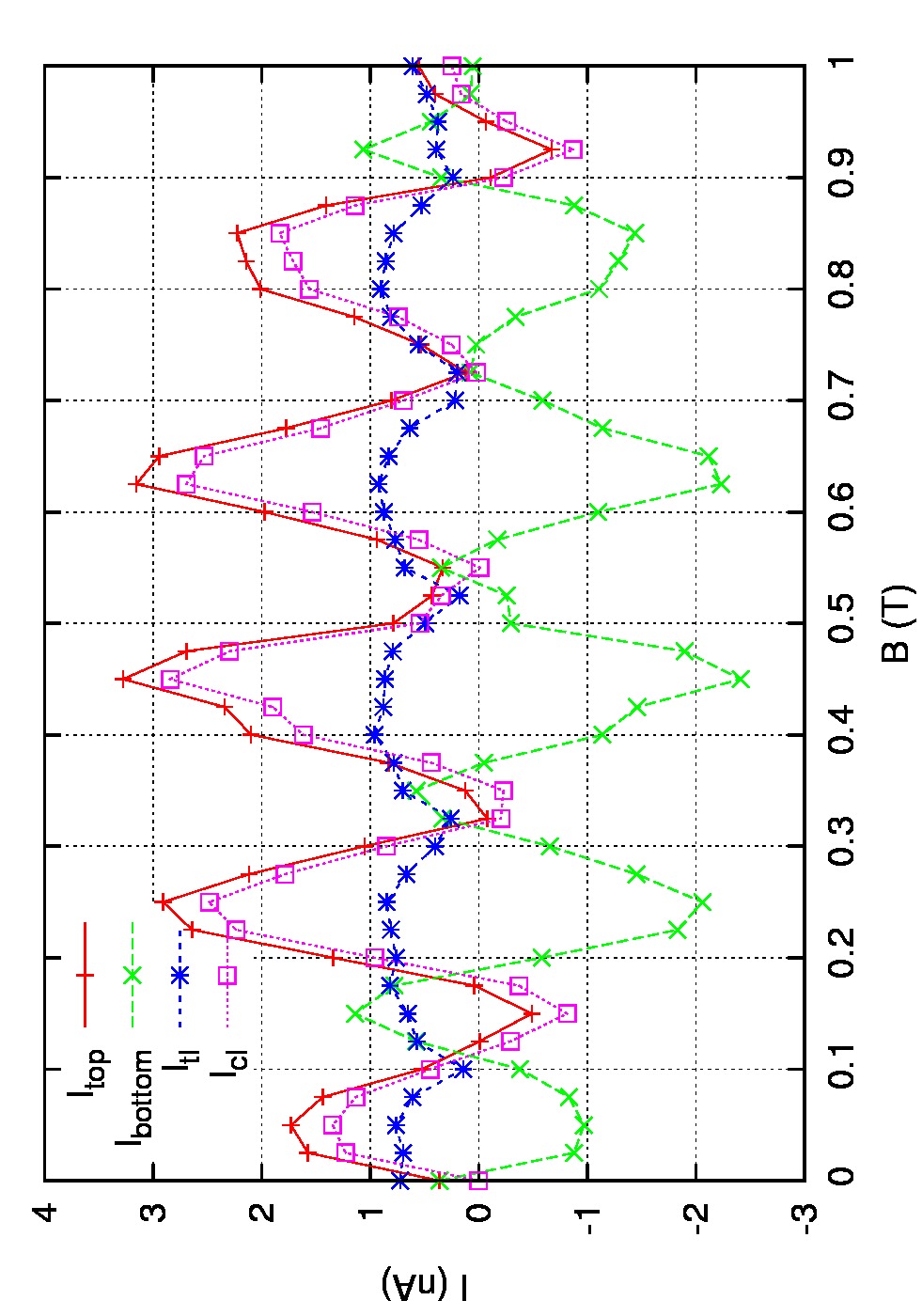}
       \caption{(Color online) Local current through the top arm of the ring $I_{\mathrm{top}}$ (solid red),
       local current through the bottom arm of the ring $I_{\mathrm{bottom}}$,
       total local current $I_{\rm tl}$,
       and the circular local current
       $I_{\rm cl}$) versus the magnetic field
       averaged over the time interval $[180,220]\mathrm{ps}$ in the case of x-polarized photon field.}
       \label{current_arm_x_200}
\end{figure}
In \fig{current_arm_x_200}, we show the magnetic field dependence of
the partial local currents $I_{\mathrm{top}}$ and
$I_{\mathrm{bottom}}$ through the top and bottom arms, the total
local current $I_{\rm tl}$ across $x=0$,  and the circular local
current $I_{\rm cl}$, which are convenient tools to study the
relative importance of local ``persistent'' current flows induced by the magnetic
field in the long-time response transient time regime. We averaged the local currents
over the time interval $[180,220]\mathrm{ps}$ around $t=200$~ps to soften possible high frequency fluctuations
(compare with \fig{current_all_x_vt}). In general, the top local
current exhibits opposite sign to the bottom local current, and
hence the circular local current (solid purple) is usually larger
in magnitude than the total local current (dotted blue). The local current
through the two current arms, $I_{\rm tl}$, is strongly suppressed
in the case of half integer flux quanta showing a very similar
behavior to the nonlocal currents $I_L$ and $I_R$
(\fig{current_per_200_x}). This is because the destructive
interference in the quantum ring enhances the back scattering 
for magnetic flux with half integer quanta.

In the absence of magnetic field $B=0$, the circular current is
identical to zero due to the symmetric situation for both ring arms.
Moreover, it is interesting to note that the circular local current
$I_{\rm cl}$ reaches $1.347$~nA for less than half a flux quantum
(at $B=0.05$~T), increases further until $B=0.45$~T with a maximum
value $\max |I_{\rm cl}|=2.844$~nA and decreases again for
$B>0.45$~T. The magnetic component of the diamagnetic part of the
circular local current increases linearly with the magnetic field
$B$, but the paramagnetic part guarantees a behavior, which is
closer to being periodic with the flux quantum. The periodic part is in analogy to the behavior for a
ring of infinitesimal width.~\cite{PhysRevLett.7.46}   In the case
of high magnetic field regime ($B>0.45$~T), a comparison with
\fig{MB_spec_vB_x} shows that the different flux periods of
different MB-states in the finite-width ring lead to destructive
interference effects reducing the periodic oscillations
considerably. The most common direction of the
circular current is clockwise. This is because the vortices at the
lead connection areas are threaded by less than half a flux quantum
leading to counter-clockwise vortex rotation direction. Then, as a
consequence of charge continuity and the vortex location outside the
ring radius, clockwise direction for $I_{\rm cl}$ is preferred.

\subsection{Photons with y-polarization}

\begin{figure}[htbq] 
       \includegraphics[width=0.46\textwidth,angle=-90]{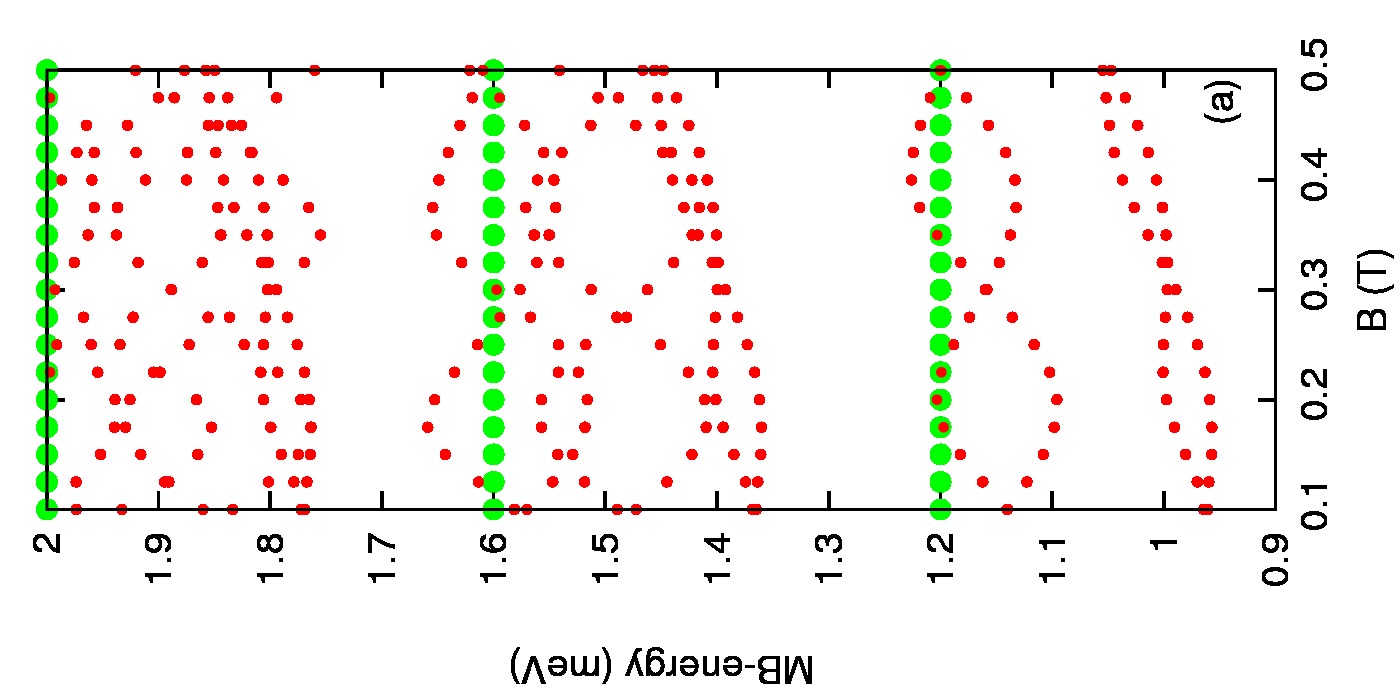}
       \includegraphics[width=0.46\textwidth,angle=-90]{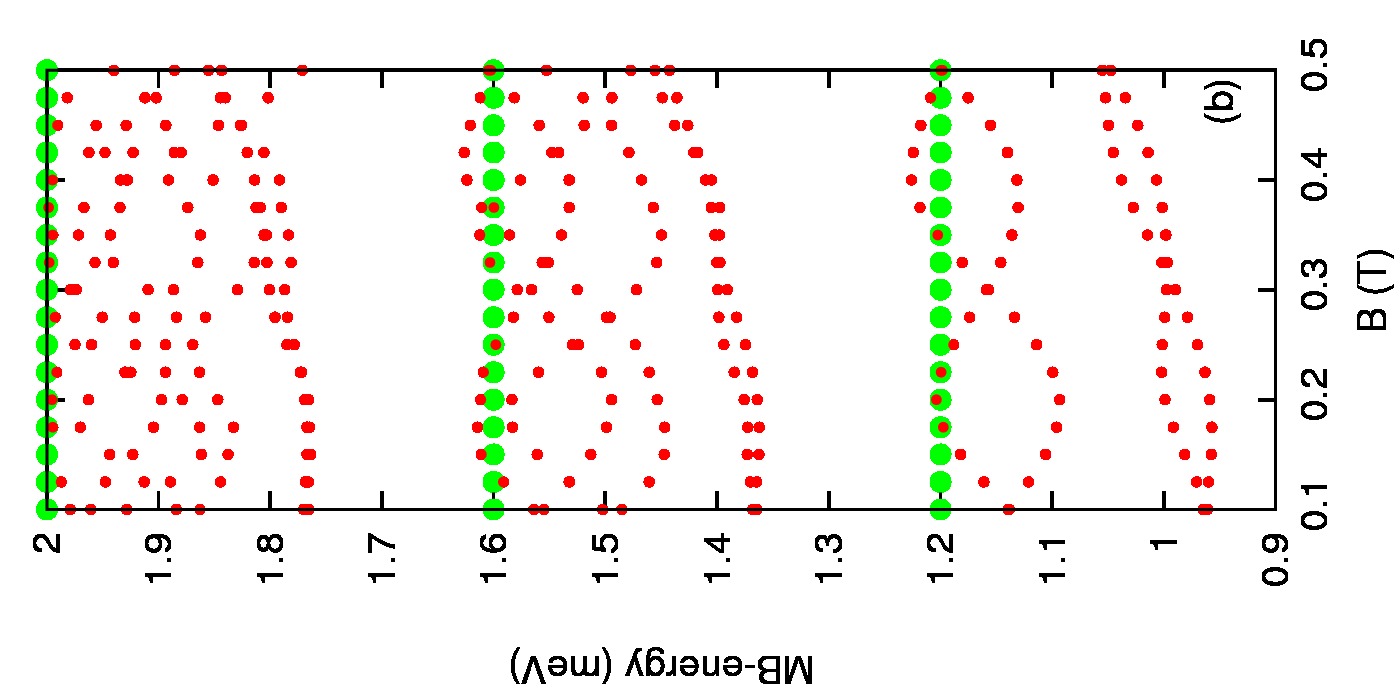}
       \caption{(Color online)
       MB energy spectrum of the system Hamiltonian $\hat{H}_{S}$ versus magnetic field
       $B$ within the bias window energy range for (a) x-polarized and (b) y-polarized photon
       field.   The states are differentiated according to their electron content $N_e$:
       zero-electron states ($N_e=0$, 0ES, green dots)
       and single electron states ($N_e=1$, 1ES, red dots).}
       \label{MB_spec_vB_red_win}
\end{figure}

In this subsection, we focus on the y-polarized photon field situation
and compare with the results for the x-polarized photon field. Figure
\ref{MB_spec_vB_red_win} shows the MB energy spectra of the system
Hamiltonian $\hat{H}_{S}$ in the case of (a) x-polarized and (b)
y-polarized photon field. We note in passing that \fig{MB_spec_vB_red_win}(a) 
magnifies a part of the MB spectrum of \fig{MB_spec_vB_x}. The mostly occupied levels are the two
levels around $1.4$~meV.  In the cases of both x- and y-polarized photon
field, we see the MB energy degeneracy around $B= 0.1$ and $0.325$~T
related to the destructive AB phase interference. However, in the
case of y-polarization, an extra MB energy degeneracy is found at
$B=0.425$~T.  This degeneracy is related to a {\em photon}
suppressed current dip, i.e. not related to AB
oscillations.

\begin{figure}[htbq]
       \includegraphics[width=0.34\textwidth,angle=-90]{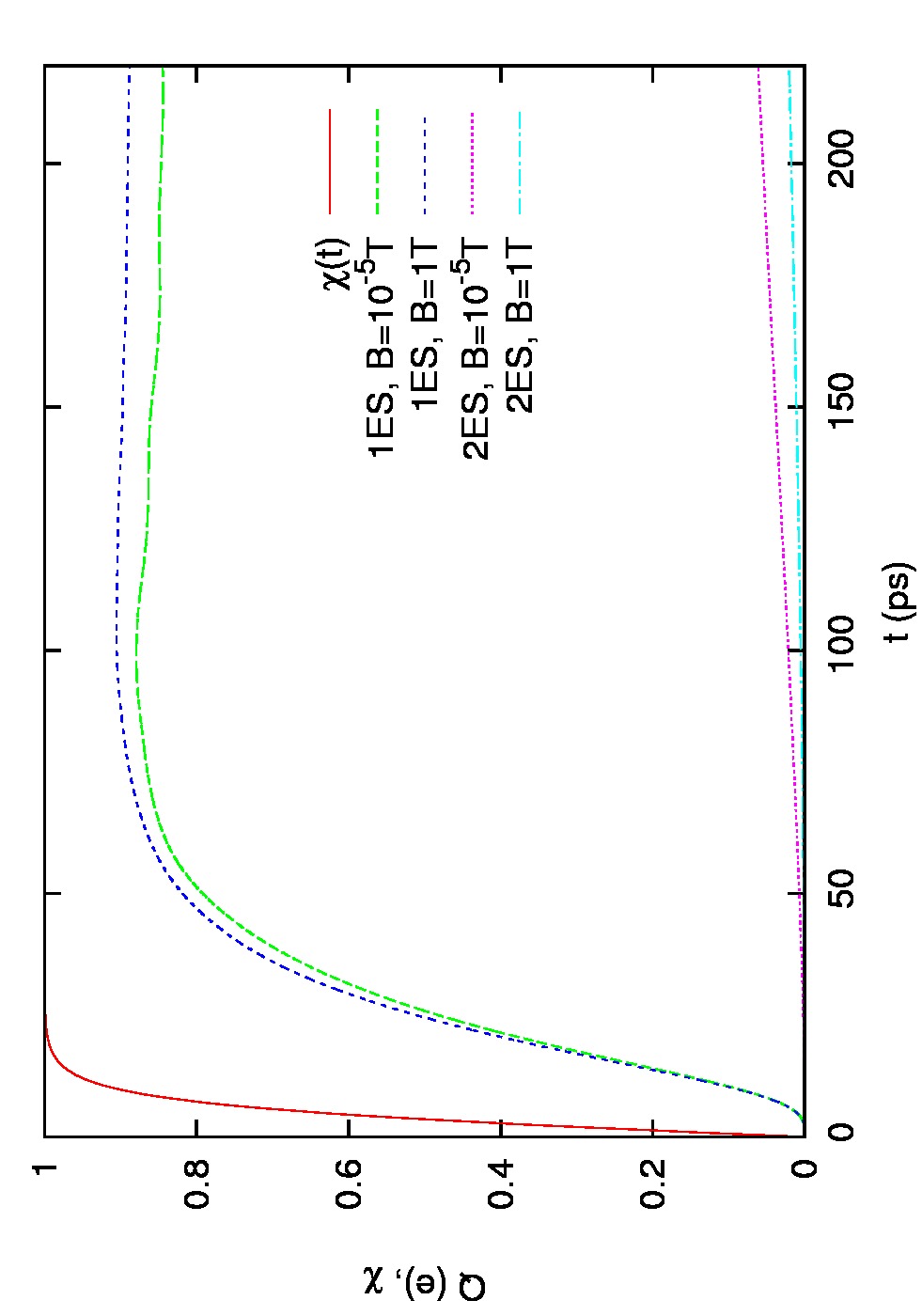}       
       \caption{(Color online) Switching function $\chi^{l}(t)$ (solid red),
       charge of all 1ES for $B=10^{-5}$~T (dashed green) and
       $B=1.0$~T (dotted blue), and charge of all 2ES for $B=10^{-5}$~T (dotted purple) and
       $B=1.0$~T (dash-dotted cyan)
       as a function of time. The photon field is y-polarized. }
       \label{1ES_2ES_y}
\end{figure}

Figure \ref{1ES_2ES_y} illustrates the charge of 1ES and 2ES as a
function of time. In the case of low magnetic field regime
$B=10^{-5}$~T, we notice that $(t,Q_{\rm 1ES}) = (200\, {\rm ps},
0.847e)$ and $(t,Q_{\rm 2ES}) = (200\, {\rm ps}, 0.055e)$.  In the
case of high magnetic field regime $B=1$~T, we notice that
$(t,Q_{\rm 1ES}) = (200\, {\rm ps}, 0.891e)$ and $(t,Q_{\rm 2ES}) =
(200\, {\rm ps}, 0.017e)$. The 2ES are occupied much slower
than the 1ES due to similar reasons than for the x-polarized photon field.  The magnetic
field enhances the 1ES occupation by $\delta Q_{\rm 1ES} =
0.044e$ while it suppress the 2ES occupation by $\delta Q_{\rm 2ES} = -0.038e$.
Therefore, in comparison with the case of x-polarization shown in
\fig{1ES_2ES_x}, we realize that
the y-polarized photon field
mildens the
influence of the applied magnetic field on the charging feature.

\begin{figure}[htbq]
       \includegraphics[width=0.34\textwidth,angle=-90]{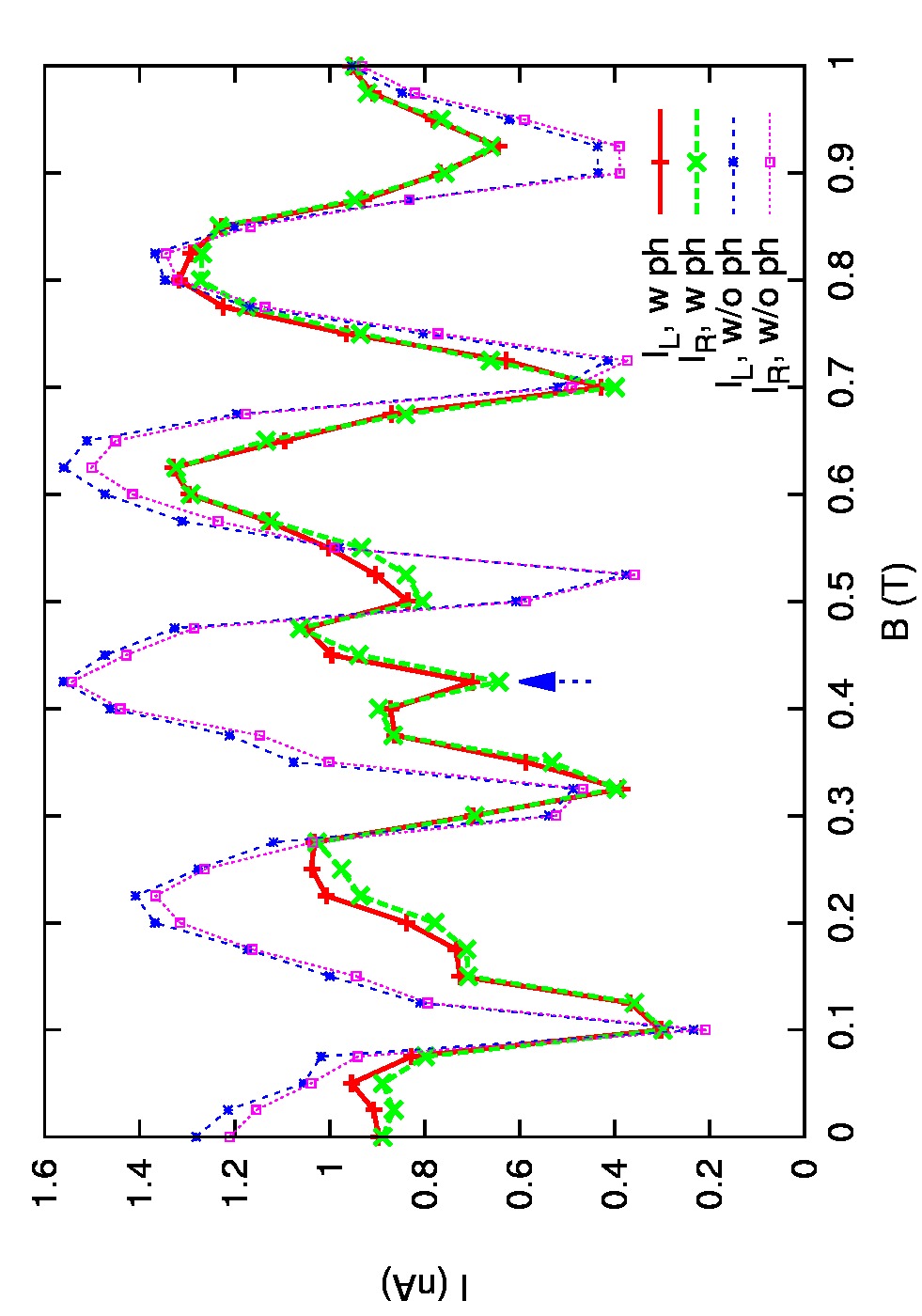}
       \caption{(Color online) Left charge current $I_{L}$ (solid red) and right charge current
       $I_R$ (long-dashed green) versus the magnetic field with (w) y-polarized photon field
       at $t=200\mathrm{ps}$. For comparison: left charge current $I_{L}$ (short-dashed blue) and right charge current
       $I_R$ (dotted purple) in a purely electronic central system, i.e. without (w/o) photon cavity.}
       \label{current_per_200_y}
\end{figure}
Figure \ref{current_per_200_y} shows the left charge current $I_{L}$
(solid red) and the right charge current $I_R$ (dashed green) as a
function of magnetic field in the case of y-polarized photon field
at $t=200\mathrm{ps}$.  The similar values of $I_L(B)$ and $I_R(B)$
agree well with the long-time response regime slow-down in charging predicted in \fig{1ES_2ES_y}, 
which is almost completed for the 1ES. Charge current oscillations
are shifted slightly from the period $B_0 \approx 0.2$~T due to the
broad ring geometry.  Moreover, the oscillation amplitude and extrema positions show more
unexpected features than in the case of x-polarized photon field.  The
first current minimum is at magnetic field $B=0.1$~T (with a half
flux quantum) with left charge current $I_L=0.303$~nA and right
charge current $I_R=0.298$~nA. At magnetic field $B=0.225$~T 
corresponding to the case of one flux quantum, the left
charge current $I_L=1.007$~nA and the right charge current
$I_R=0.935$~nA. It is interesting to point out that the magnetic
field dependence of the charge current exhibits a pronounced dip at
magnetic field $B=0.425$~T (two flux quanta) in the case of
y-polarized photon field that is not present in the case of
x-polarized photon field.

The dip structure in the charge current at $B=0.425$~T
is due to the above mentioned degeneracy of the MB energy spectrum, which strongly
suppresses the photon-assisted tunneling feature.
Furthermore, the charge current can be enhanced by the
y-polarized photon field at magnetic field with half integer flux
quantum: the y-polarized photon field significantly influences the quantum
interference of the circular local current flow including the 
destructive interference feature of the charge current.

\begin{figure}[htbq]
       \subfigure[]{
       \includegraphics[width=0.34\textwidth,angle=-90]{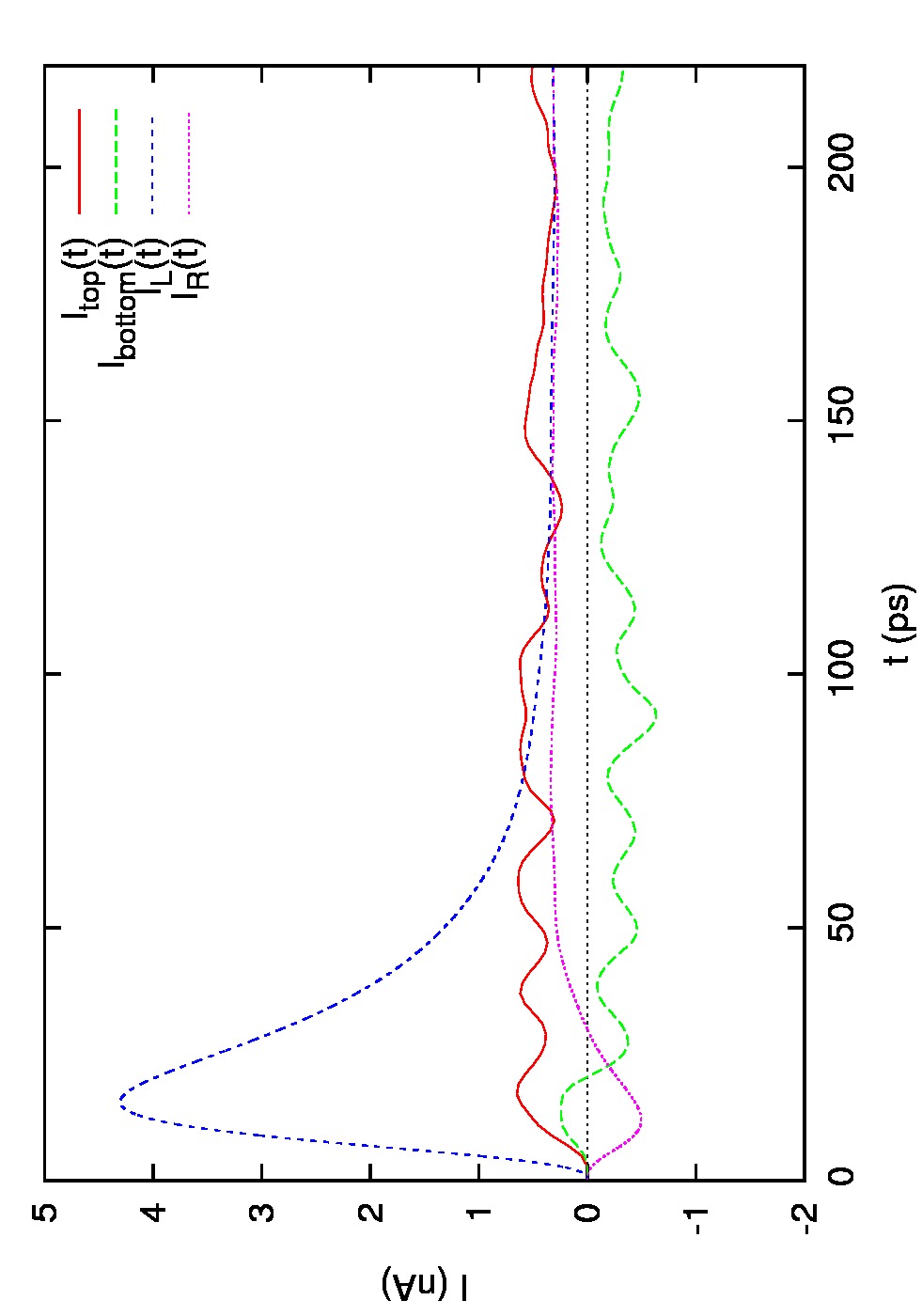}
       \label{current_all_y_vt_B010}}
       \subfigure[]{
       \includegraphics[width=0.34\textwidth,angle=-90]{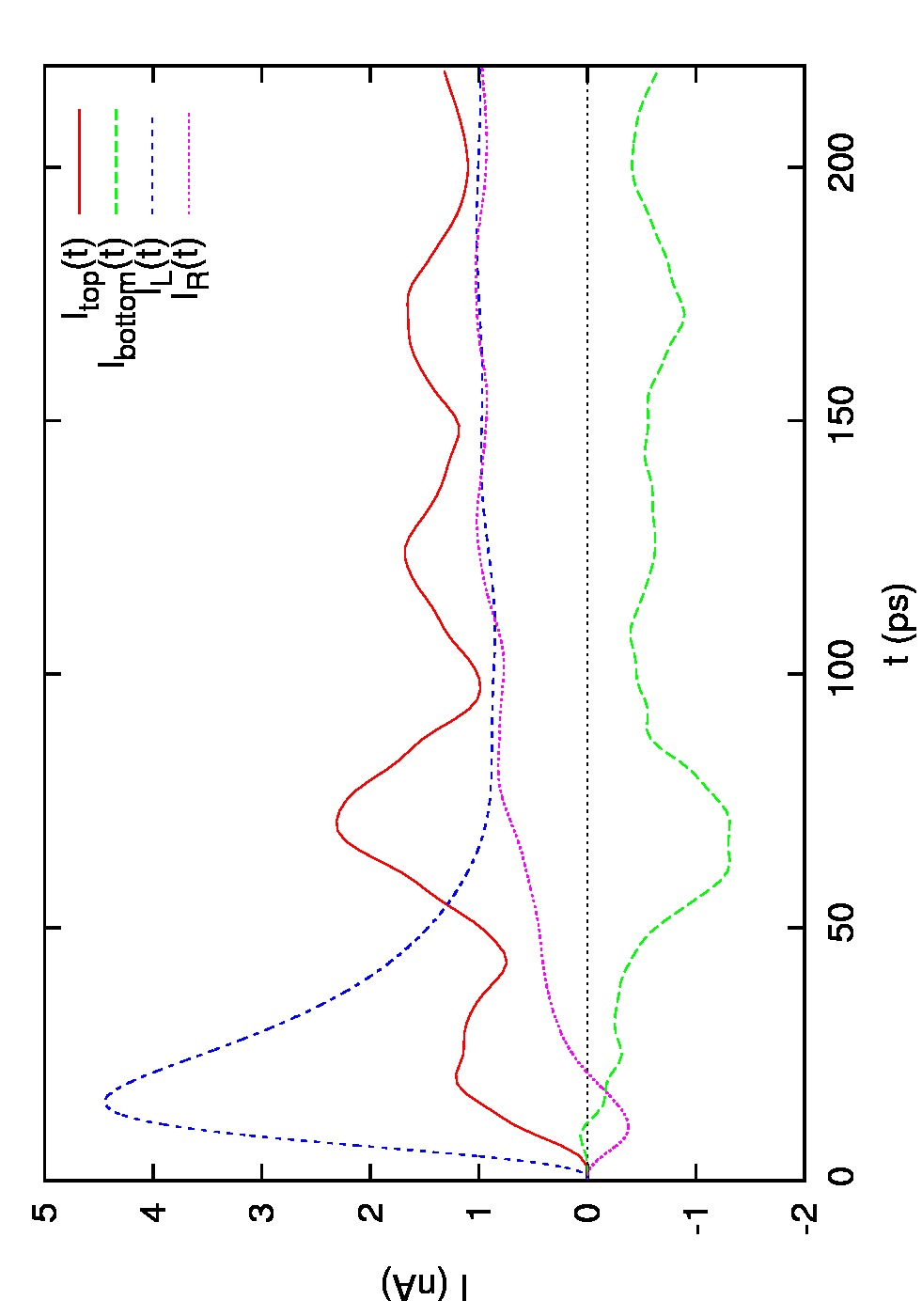}
       \label{current_all_y_vt_B0225}}
       \caption{(Color online) Local current through the top ring branch ($I_{\mathrm{top}}(t)$)
       and bottom ring branch ($I_{\mathrm{bottom}}(t)$), current from the left lead ($I_{L}(t)$)
       and into the right lead ($I_R(t)$) for \subref{current_all_y_vt_B010}
       $B=0.1$~T
       and \subref{current_all_y_vt_B0225} $B=0.225$~T in the case of y-polarized photon field.}
       \label{current_all_y_vt}
\end{figure}
Figure \ref{current_all_y_vt} illustrates the time-evolution of the
left total current $I_L(t)$, the right total current $I_R(t)$, the
top local current $I_{\mathrm{top}}(t)$, and the bottom local
current $I_{\mathrm{bottom}}(t)$.  In the case of $B=0.1$~T shown in
\fig{current_all_y_vt}(a), the maximum value of the current from the
left lead into the system at $t=15.55$~ps is $I_L(t)=4.303$~nA.
Furthermore, the minimum value of the charge current into the right lead at
$t=12.00$~ps is $I_R = -0.494$~nA. The negative right charge current
indicates that the central system is charged from the left and the
right lead for a short time.  This charging from the right is a little
weaker than in the case of x-polarized photon field. In the case of
$B=0.225$~T shown in \fig{current_all_x_vt}(b), the maximum value of
the left current at $t=15.54$~ps is $I_L=4.446$~nA
and the minimum value of the right current at $t=10.80$~ps is
$I_R=-0.337$~nA. Hence, the integer magnetic
flux enhances the charge accumulation from the left lead, while it suppresses the short-time
regime charging from the right lead assisting the net current flow from the left to the right 
already in the highly non-equilibrium situation in the very beginning.
The local charge transport may differ in direction in the ring arms 
due to the ``persistent'' current induced by the magnetic field.

\begin{figure}[htbq]
       \includegraphics[width=0.37\textwidth,angle=0,viewport=40 8 218 170,clip]{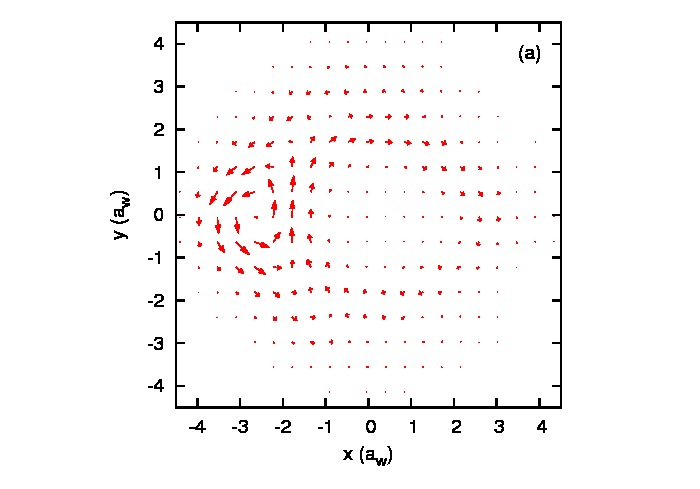}
       \includegraphics[width=0.37\textwidth,angle=0,viewport=40 8 218 170,clip]{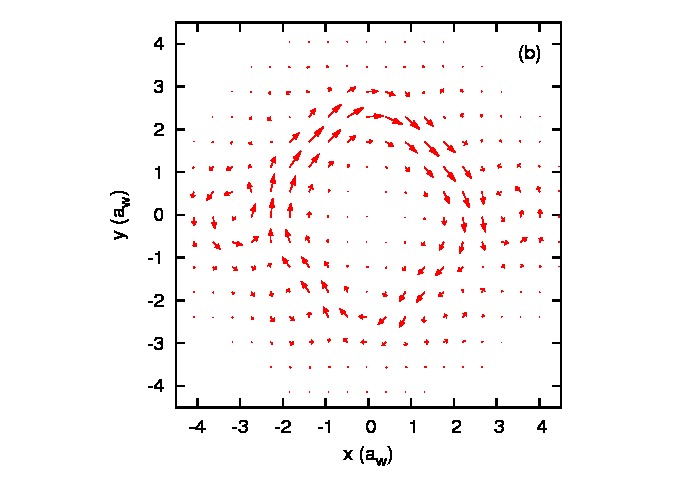}
       \includegraphics[width=0.37\textwidth,angle=0,viewport=40 8 218 170,clip]{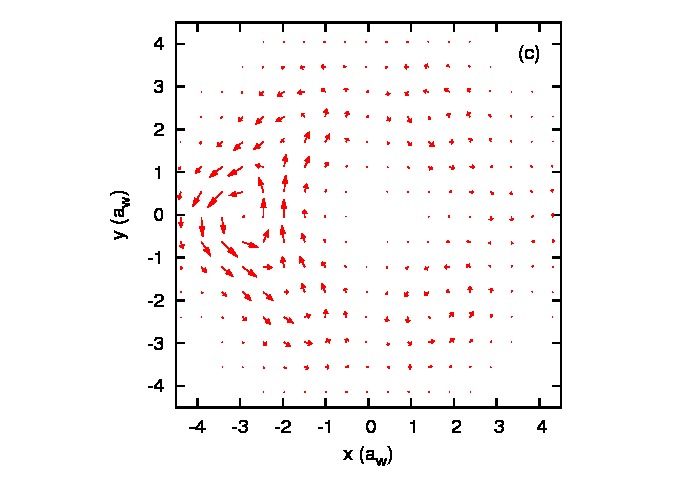}
       \caption{(Color online) Normalized charge current density vector field in the central system
       for (a) $B=0.1$~T, (b) $B=0.225$~T and
       (c) $B=0.425$~T at $t=200$~ps
       in the case of y-polarized photon field.}
       \label{cur_den_y_200}
\end{figure}

In \fig{cur_den_y_200}, we demonstrate the normalized charge current
density vector field $\mathbf{j}(\mathbf{r},t)$ in the central ring
system for the magnetic field, (a) $B=0.1$~T, (b) $B=0.225$~T, and
(c) $B=0.425$~T, in the long-time response regime $t=200$~ps in the
case of y-polarized photon field. For magnetic field $B=0.1$~T, a
clear counter-clockwise vortex can be found being associated with a long-living
localized state which is strongly dominating the current flow pattern in the
central ring system, as is shown in \fig{cur_den_y_200}(a). However, for magnetic field
$B=0.225$~T, this counter-clockwise vortex appears weaker relative to the total local current, 
but is present at both contact regions as shown in \fig{cur_den_y_200}(b). 
Figures \ref{cur_den_y_200}(a) and \ref{cur_den_y_200}(b) are similar to
Figs. \ref{cur_den_x_200}(a) and \ref{cur_den_x_200}(b) meaning that the local current flow
is mainly governed by AB interference with the photon polarization
having only a minor effect. 

Figure \ref{cur_den_y_200}(c) shows the current
density for $B=0.425$~T (two flux quanta), which is similar to
\fig{cur_den_y_200}(a) (a half flux quantum) and not to the one flux quantum case 
as for x-polarization (similarity between \fig{cur_den_x_200}(c) and \fig{cur_den_x_200}(b)).
This feature is not predicted by the AB effect, but is caused by the influence of the y-polarized photons.
However, the impact of a MB spectrum degeneracy of the mostly occupied MB states (\fig{MB_spec_vB_red_win}(b))
 on the local current flow structure is similar 
whether the degeneracy is in agreement with the AB effect (\fig{cur_den_y_200}(a)) or not, 
i.e. originates from the photons (\fig{cur_den_y_200}(c)).

\begin{figure}[htbq]
       \subfigure[]{
       \includegraphics[width=0.34\textwidth,angle=-90]{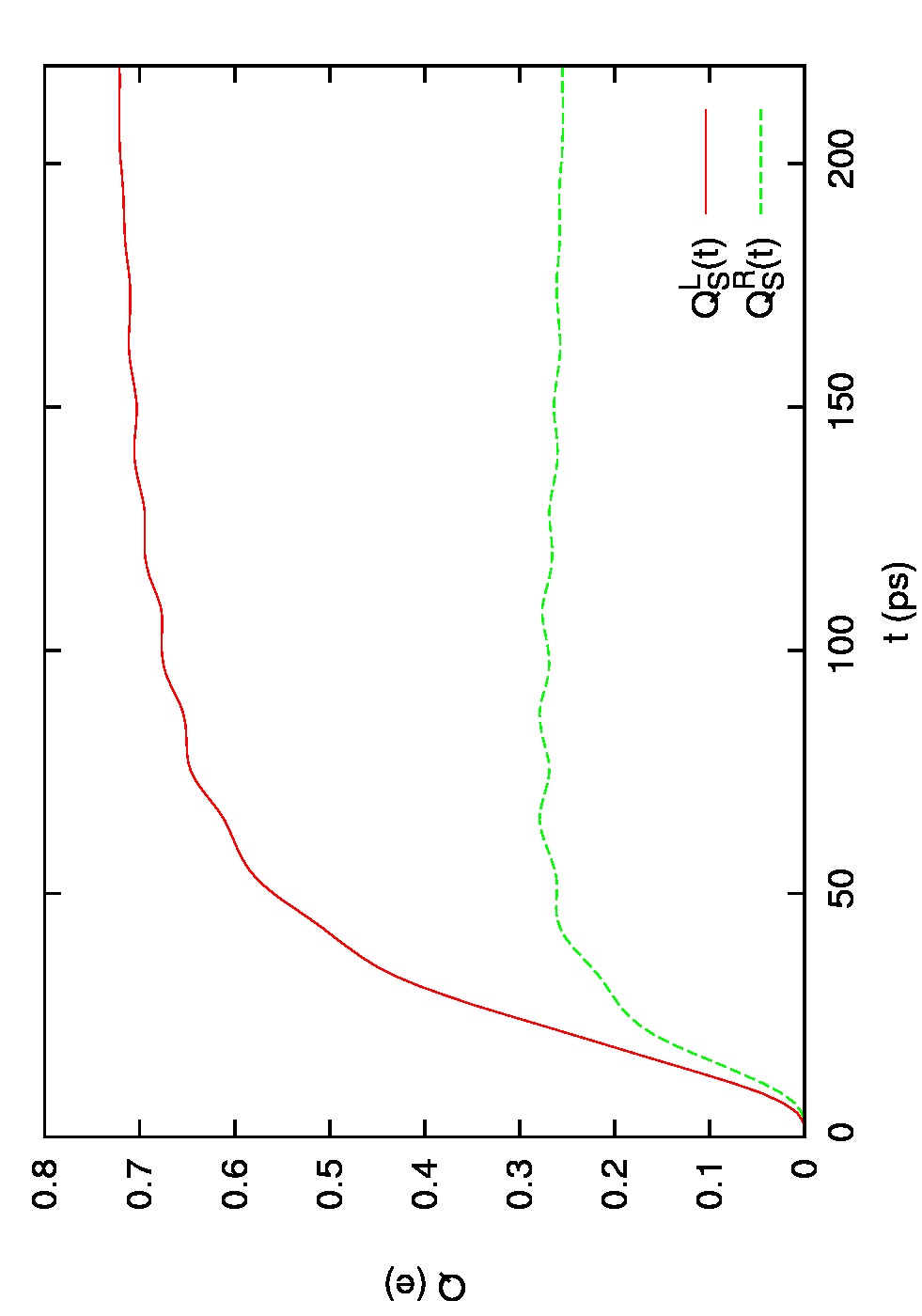}
       \label{density_all_y_vt_B010}}
       \subfigure[]{
       \includegraphics[width=0.34\textwidth,angle=-90]{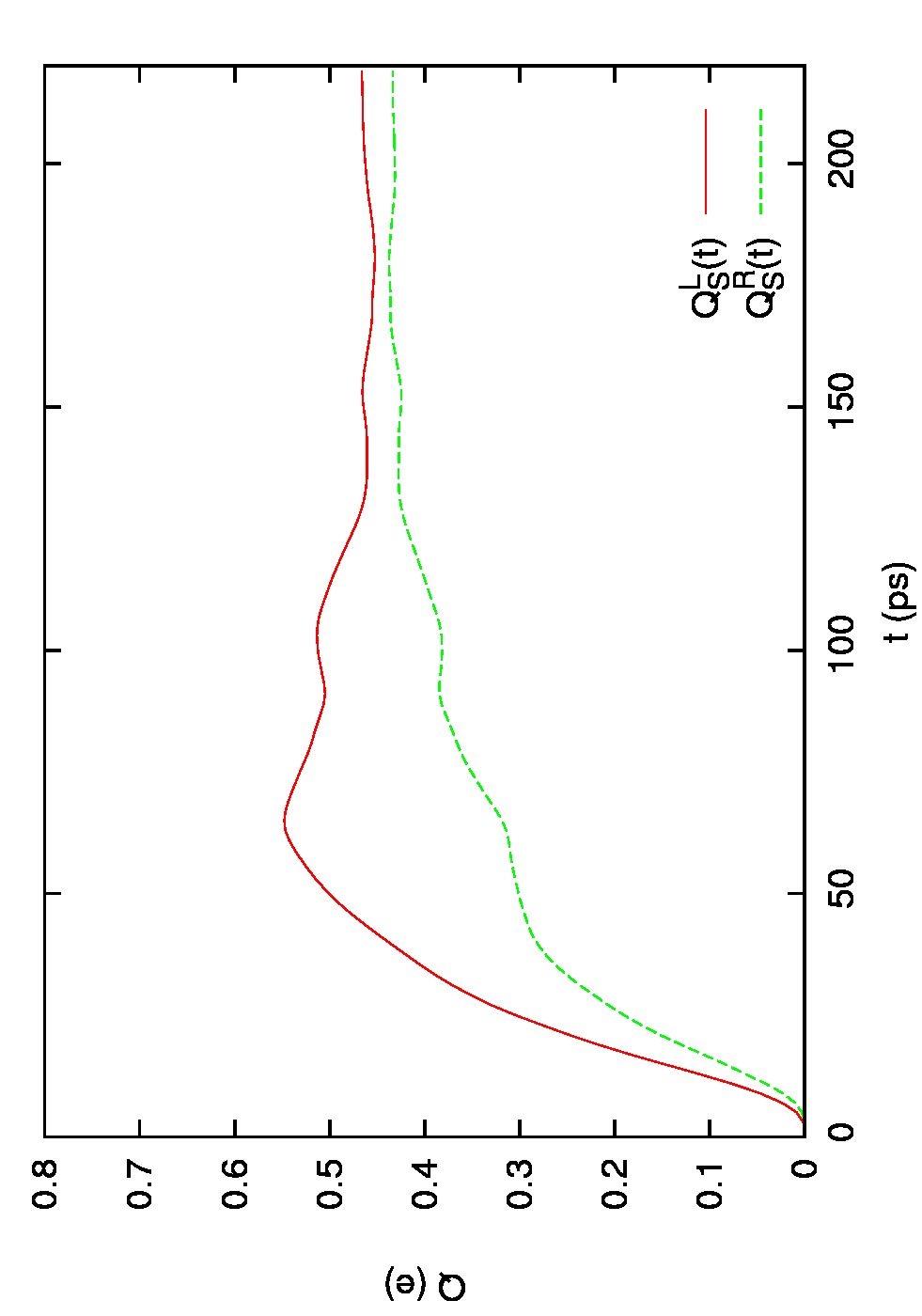}
       \label{density_all_y_vt_B0225}}
       \caption{(Color online) Charge in the left half ($Q_S^{L}(t)$) or right half ($Q_S^{R}(t)$)
       of the central system as a function of time
       for \subref{density_all_y_vt_B010} $B=0.1$~T and \subref{density_all_y_vt_B0225}
       $B=0.225$~T. The photon field is y-polarized.}
       \label{density_all_y_vt}
\end{figure}

Figure \ref{density_all_y_vt} shows the time-evolution of
$Q_S^{L}(t)$ and $Q_S^{R}(t)$. In the short-time response regime at $t=65$~ps, the charge on the
left and right part of the ring are $Q_S^{L}=0.548e$ and
$Q_S^{R}=0.318e$ for $B=0.225$~T, respectively; the charge on the
left and right part of the ring are $Q_S^{L}=0.610e$ and
$Q_S^{R}=0.279e$ for $B=0.1$~T. Consequently, the
electron dwell time on the left-hand side of the ring is enhanced
while the electron dwell time on the right-hand side is suppressed
for y-polarized photon field and integer flux quantum; 
this feature is though more pronounced for a half flux quantum 
and y-polarized photon field, but was already described for x-polarized photons (\fig{density_all_x_vt}(a)).
The reason for the difference
in left and right dwell time in the integer flux quantum case is a low
frequency oscillation of the most important close-in-energy levels.
In the long-time response regime at time $t=200$~ps, the picture is very similar to
the x-polarized photon field case: for
$B=0.225$~T, the left and right charges are of similar magnitude, $Q_S^{L}=0.462e$ and $Q_S^{R}=0.431e$;
and for $B=0.1$~T, the charge is mainly accumulated at the left hand side, $Q_S^{L}=0.720e$ and $Q_S^{R}=0.256e$.

In the $B=0.225$~T case, the MB
energies of the mostly occupied MB levels are $E_{10}^{y}=1.3846$~meV
and $E_{9}^{y}=1.3683$~meV such that $\Delta E_{9,10}^{y}=0.0163$~meV. The
energy level difference of the mostly occupied MB levels is only
$44\%$ of the case of x-polarized photon field:
$E_{9,10}^{y}\approx 0.44 \times E_{9,10}^{x}$. The corresponding TL
oscillation period of the closed system would be $\tau_{\rm
TL}^{0}=254$~ps.  The oscillation period is too long to be observed
clearly in \fig{density_all_y_vt}\subref{density_all_y_vt_B0225},
but the first maximum of $Q_S^{L}(t)$ at $t= 65$~ps represents the
starting point of the low frequency oscillation, which is better
visible in the TL system defined by the two mostly occupied states.
Our findings suggest that the energy difference of the two mostly
occupied levels controls not only the charge distribution, but also
photonic suppressions of the AB current. The different connectivity
(probability density on the left or right ring part) to the leads
found within the TL dynamic suggests that the probability of a
photon coupled electron transition between these levels plays a
major role in understanding the photonic modifications of the AB
current pattern.

\begin{figure}[htbq]
       \subfigure[]{
       \includegraphics[width=0.37\textwidth,angle=0,viewport=45 12 240 152,clip]{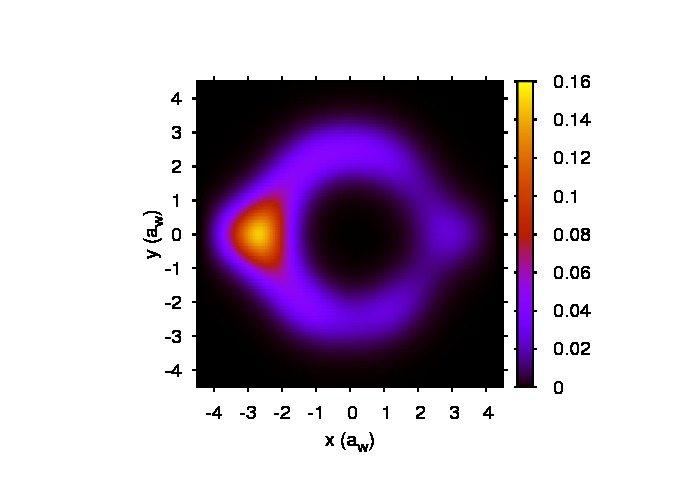}
       \label{den_B010y_200}}
       \subfigure[]{
       \includegraphics[width=0.37\textwidth,angle=0,viewport=45 12 240 152,clip]{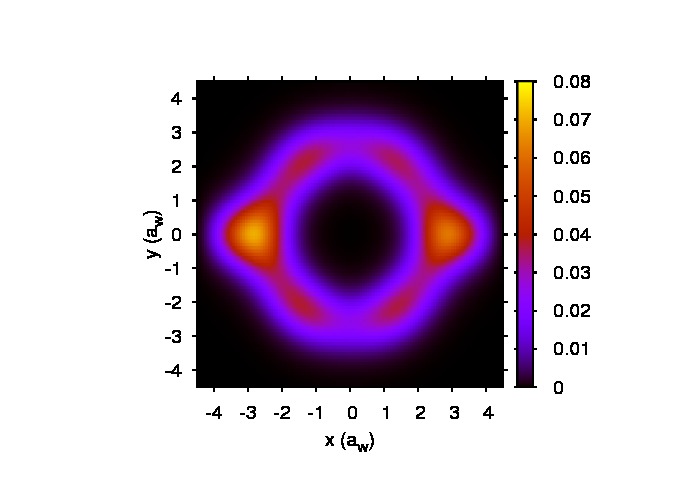}
       \label{den_B0225y_200}}
       \subfigure[]{
       \includegraphics[width=0.37\textwidth,angle=0,viewport=45 12 240 152,clip]{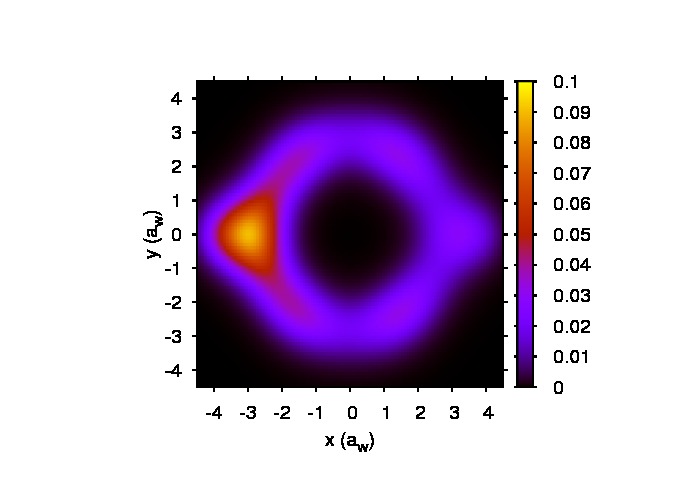}
       \label{den_B0425y_200}}
       \caption{(Color online) Charge density distribution $\rho(\mathbf{r},t)\; (e/a_w^2)$ in the central system
       for \subref{den_B010y_200} $B=0.1$~T, \subref{den_B0225y_200} $B=0.225$~T, and
       \subref{den_B0425y_200} $B=0.425$~T in the y-polarized photon field case at $t=200$~ps.}
       \label{den_y_200}
\end{figure}

Figure \ref{den_y_200} shows the charge density distribution in the
central ring system in the case of y-polarized photon field for
magnetic field \subref{den_B010y_200} $B=0.1$~T,
\subref{den_B0225y_200} $B=0.225$~T, and \subref{den_B0425y_200}
$B=0.425$~T at $t=200$~ps. In the case of $B=0.1$~T shown in
\fig{den_y_200}\subref{den_B010y_200}, the electrons are highly
accumulated on the left-hand side of the quantum ring with very weak
coupling to the right lead, and hence strongly blocking the left
charge current and suppressing the right charge current. 

In the case
of $B=0.225$~T shown in \fig{den_y_200}\subref{den_B0225y_200}, the
electrons manifest oscillating feature between the left and right
end of the quantum ring. At time $t=200$~ps, the electrons are
equally well accumulated on both sides of the quantum ring.
This situation is related to the manifestation of the current peaks
observed in \fig{current_per_200_y}. The
charge is redistributed when compared to magnetic field
$B=0.1$~T. We observe a depletion of about $50\%$ on the left-hand
side with equivalent charge augmentation on the right hand side. The
magnetic field $B=0.225$~T with integer flux quanta enhances the
likelihood for electrons to flow through the quantum ring to the
right-hand side of the central system and further to the right lead.

Figure \ref{den_y_200}\subref{den_B0425y_200} shows the
charge density for $B=0.425$~T (two flux quanta), which is similar
to \fig{den_y_200}\subref{den_B010y_200} (a half flux quantum) and not to the one flux quantum case 
as for x-polarization (similarity between \fig{den_B0425x_200} and \fig{den_B0225x_200}).
This feature is not predicted by the AB effect, but is caused by the influence of the y-polarized photons.
However, the impact of a MB spectrum degeneracy of the mostly occupied MB states (\fig{MB_spec_vB_red_win}(b))
 on the density distribution is similar 
whether the degeneracy is in agreement with the AB effect (\fig{den_B010y_200}) or not, 
i.e. originates from the photons (\fig{den_B0425y_200}).

\begin{figure}[htbq] 
       \includegraphics[width=0.34\textwidth,angle=-90]{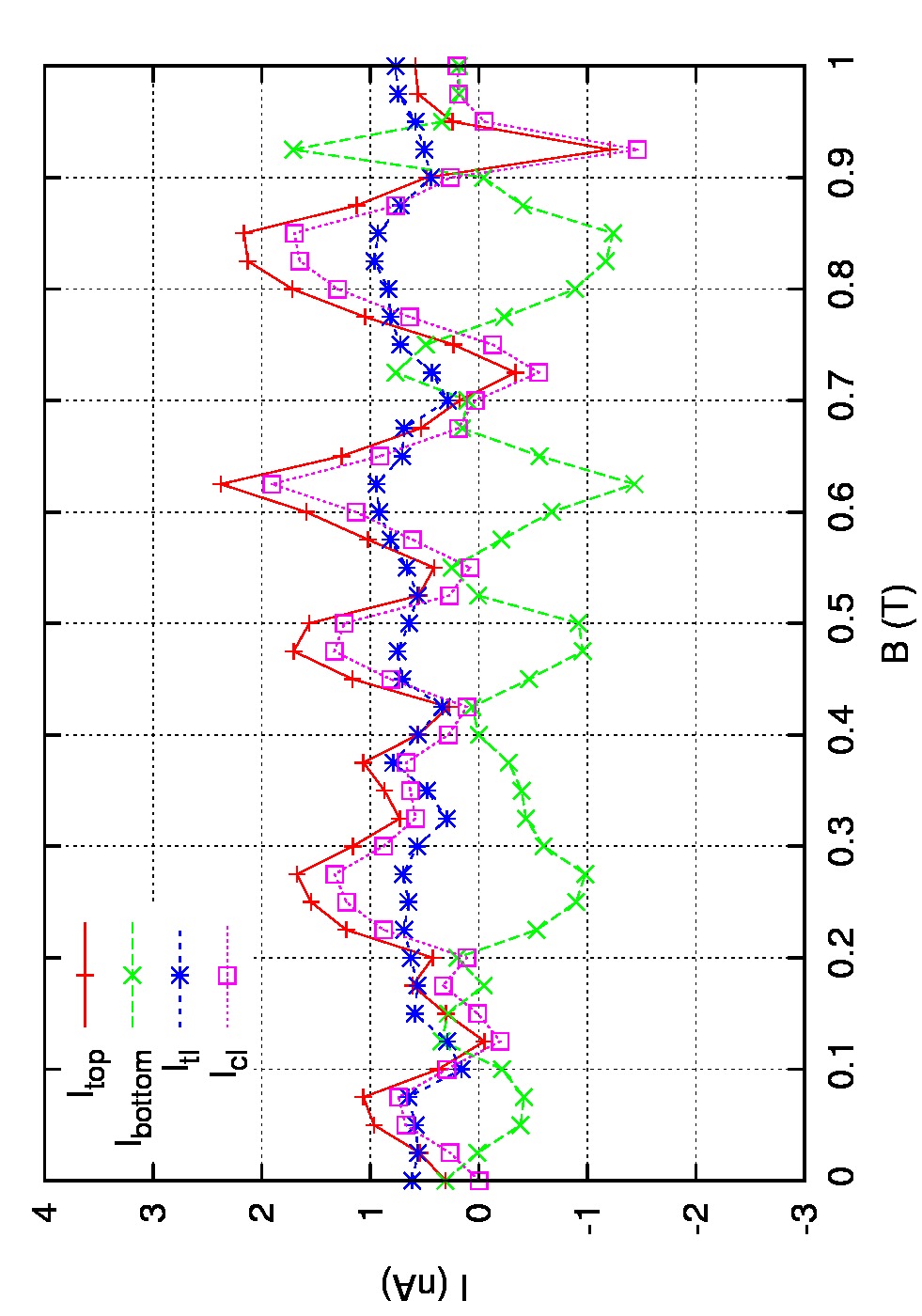}
       \caption{(Color online) Local current through the top ring arm ($I_{\mathrm{top}}$)
       and bottom ring arm ($I_{\mathrm{bottom}}$) and total local current
       ($I_{\rm tl}$) and circular local current
       ($I_{\rm cl}$) versus the magnetic field and
       averaged over the time interval $[180,220]\mathrm{ps}$
       in the case of y-polarized photon field.}
       \label{current_arm_y_200}
\end{figure}

In \fig{current_arm_y_200}, we show the magnetic field dependence of
the local currents $I_{\mathrm{top}}$ and
$I_{\mathrm{bottom}}$ through the top and bottom arms, respectively, the total
local current $I_{\rm tl}$ across $x=0$,  and the circular local
current $I_{\rm cl}$. We averaged the local currents over the time
interval $[180,220]\mathrm{ps}$ around
$t=200$~ps to soften possible high frequency fluctuations (compare with
\fig{current_all_y_vt}). In most cases (however less often than for x-polarized photons),
the top local current exhibits opposite sign to the bottom local current, and hence the circular
local current (solid purple) is usually larger than the total local
current (dotted blue). The local current through the two current
arms, $I_{\rm tl}$, is suppressed in the case of half integer flux
quanta showing a similar behavior to the nonlocal currents $I_L$ and
$I_R$ (\fig{current_per_200_y}), but with more irregularities
 due to the stronger effective influence of the y-polarized photon field. It
is interesting to note that the current suppression dip at
$B=0.425$~T (marked by the blue arrow in \fig{current_per_200_y})
appears also in the local current (blue dashed curve) flowing
through both ring arms from the left to the right.

The circular local current reaches a maximum absolute value of $\max
|I_{\rm cl}| =1.905$~nA at $B=0.625$~T, which is by $0.939$~nA
smaller than for x-polarization. It is clearly visible
from a comparison of \fig{current_arm_y_200} and
\fig{current_arm_x_200} that the circular current is considerably
smaller than in the x-polarized photon case, while the total local
current is of the same order. Thus, the capability of the magnetic
field to drive a rotational ring current is weakened by having the
electromagnetic field y-polarized. In particular, this can be said
about the diamagnetic part of the circular local current leading to
the much smaller value $I_{\rm cl} =0.675$~nA at low magnetic field
$B=0.05$~T. The periodicity of the circular local current is
preserved better for x-polarized photon field as is for the total
local current. We note here in passing that it is not possible to
understand the non-trivial magnetic field dependence of $I_{\rm cl}$ by
resorting solely to a TL description.

\section{Concluding Remarks}

We have presented a time-convolutionless generalized master equation
formalism that allows us to calculate the non-equilibrium transport
of Coulomb interacting electrons through a broad quantum ring in a
photon cavity under the influence of a uniform perpendicular
magnetic field.  The topologically nontrivial broad ring geometry
allows for substantial electron-electron correlations relative to their
kinetic energy and, hence, a large basis is required for
sufficient numerical accuracy.  The magnetic field, however,
increases slightly the energy difference of the 2ES with respect to
the 1ES, thus enhancing the 1ES occupation while suppressing the 2ES
occupation. The central quantum ring 1ES are charged quickly from both leads.
Electron-electron correlation and sequential tunneling slow down the 2ES charging in the
long-time response regime. Aharonov-Bohm charge current oscillations
can be recognized in the long-time response regime with magnetic field period $B_0= \Phi_0 /A$, which is related to the flux
quantum $\Phi_0$ and ring area $A$.

In the case of x-polarized photon field, we
have found charge oscillations between the left and right part of the
quantum ring when the magnetic field is associated with integer flux
quanta. The oscillation frequency agrees well with the energy
difference of the two mostly occupied states. The relatively high
energy difference for x-polarized photons is related to a relatively
high transient current through the ring. The amplitude of the charge oscillations
through the quantum ring is decreasing in
time due to dissipation effects caused by the coupling to the leads.
In general, the local current through the upper ring arm exhibits opposite sign to
the local current through the lower ring arm. Hence, the ``persistent'' circular local current is
usually larger than the total local charge current through both ring
arms from the left to the right. The
persistent current shows a periodic behavior with magnetic field, but with a tendency to clockwise
rotation due to the contact region vortex structure.

In the case of y-polarized photon field, the magnetic field
dependence of the left and right charge current exhibits a
pronounced dip at magnetic field $B=0.425$~T corresponding to two
flux quanta that is therefore clearly not related to the Aharonov-Bohm
effect. The dip is associated with a degeneracy of the two
mostly occupied 1ES at magnetic field associated with two flux
quanta. The additional level crossing appears only for y-polarized
photons. The generally lower energy difference of the two mostly
occupied MB states in the case of y-polarization disturbs the
constructive phase interference condition for the bias driven charge
flow through the quantum device and decreases the persistent current
magnitude.

In conclusion, we have demonstrated for our ring geometry that
y-polarized photons disturb our system stronger than x-polarized
photons, suppressing magnetic field induced currents and perturbing
flux periodicity beyond finite width effects by enhancing or suppressing bias-driven currents. 
It is interesting to compare these findings
to the quantum wire case, 
where it was found that mainly x-polarized photons attenuate the central system charging
due to a closer agreement of the photon mode energy and the characteristic electronic excitation energy
in x-direction.~\cite{Gudmundsson85:075306} 
In this paper, we have considered a more complex geometry, which 
reduces effectively the y-confinement energy $\hbar \Omega_0=1.0$~meV. 
The characteristic electronic excitation energy in y-direction may therefore be much closer to the photon mode energy $\hbar \omega =0.4$~meV, 
thus leading to a relatively strong influence of the y-polarized photon field on the electronic transport.
The conceived magnetic field influenced quantum ring system in a photon cavity could serve
as an elementary quantum device for optoelectronic applications and quantum information processing with unique
characteristics by controlling the applied magnetic field and the
polarization of the photon field.

%
\begin{acknowledgments}
      The authors acknowledge discussions of the manuscript with Olafur Jonasson.
      This work was financially supported by the Icelandic Research
      and Instruments Funds, the Research Fund of the University of Iceland, and the
      National Science Council of Taiwan under contract
      No.\ NSC100-2112-M-239-001-MY3.
\end{acknowledgments}
%
%
%
\bibliographystyle{apsrev4-1}
%

%
\end{document}